\documentclass[sn-basic]{sn-jnl}

\usepackage{graphicx}%
\usepackage{multirow}%
\usepackage{amsmath,amssymb,amsfonts}%
\usepackage{amsthm}%
\usepackage{mathrsfs}%
\usepackage[title]{appendix}%
\usepackage{xcolor}%
\usepackage{textcomp}%
\usepackage{manyfoot}%
\usepackage{booktabs}%
\usepackage{algorithm}%
\usepackage{algorithmicx}%
\usepackage{algpseudocode}%
\usepackage{listings}%
\usepackage{lineno}

\theoremstyle{thmstyleone}%
\newtheorem{theorem}{Theorem}
\newtheorem{proposition}[theorem]{Proposition}%

\theoremstyle{thmstyletwo}%

\theoremstyle{thmstylethree}%

\begin{document}

\title[Aggregation-diffusion in heterogeneous environments]{Aggregation-diffusion in heterogeneous environments}


\author{\fnm{Jonathan R.} \sur{Potts}}\email{j.potts@sheffield.ac.uk}

\affil{\orgdiv{School of Mathematical and Physical Sciences}, \orgname{The University of Sheffield}, \orgaddress{\street{Hounsfield Road}, \city{Sheffield}, \postcode{S3 7RH}, \country{UK}}}


\abstract{Aggregation-diffusion equations are foundational tools for modelling biological aggregations. Their principal use is to link the collective movement mechanisms of organisms to their emergent space use patterns in a concrete mathematical way. However, most existing studies do not account for the effect of the underlying environment on organism movement. In reality, the environment is often a key determinant of emergent space use  patterns, albeit in combination with collective aspects of motion. This work studies aggregation-diffusion equations in a heterogeneous environment in one spatial dimension. Under certain assumptions, it is possible to find exact analytic expressions for the steady-state solutions when diffusion is quadratic. Minimising the associated energy functional across these solutions provides a rapid way of determining the likely emergent space use pattern, which can be verified via numerical simulations. This energy-minimisation procedure is applied to a simple test case, where the environment consists of a single clump of attractive resources. Here, self-attraction and resource-attraction combine to shape the emergent aggregation. Two counter-intuitive findings emerge  from these analytic results: (a) a non-monotonic dependence of clump width on the aggregation width, (b) a positive correlation between self-attraction strength and aggregation width when the resource attraction is strong. These are verified through numerical simulations. Overall, the study shows rigorously how environment and collective behaviour combine to shape organism space use, sometimes in counter-intuitive ways.}




\maketitle

\section{Introduction}\label{sec:intro}

Aggregation phenomena are widespread in the natural world, from swarming \citep{roussi2020}, flocking \citep{papadopoulou2022self}, and herding \citep{stears2020mixed} of animals to cellular aggregations in embryonic patterns \citep{widelitz2003molecular} and slime mould slugs \citep{bonner2009}. This has led to a proliferation of research into the possible mechanisms that could cause these aggregations to form \citep{painter2024biological}. A popular mathematical tool for analysing this problem is the aggregation-diffusion equation \citep{carrillo2019aggregation}. This is a partial differential equation (PDE) model that assumes organisms have two aspects to their movement. One is an attraction to other nearby organisms of the same kind, often called self-attraction (where `self' refers here to the population rather than the individual), which is encoded in a non-local advection term. The other is a diffusive aspect to their movement, which is a simple catch-all for all the aspects of movement that are not explicitly related to aggregation, for example foraging or exploring.

Although models of diffusion and non-local advection have been instrumental in understanding the mechanisms of biological aggregation and related phenomena, they are typically analysed in a homogeneous environment \citep{painter2024biological, wang2023open}. This implicitly assumes that the environment in which the organisms live has negligible effect on the aggregation.  However, it is well-known that, in many biological situations, there are environmental drivers factors that  work alongside self-attraction to drive the emergent spatial patterns of organisms \citep{bastille2018applying, hueschen2023wildebeest,  morales2021embryos, strandburg2017habitat}. 

For example, in the embryonic development of hair and feather follicles, cells form aggregates that are driven at least in part by movement up the gradient of a chemical attractant from a point source. They may also have self-aggregation properties to their movement. Currently, it is often not clear which is the principal driver of this movement, or whether both aspects work in combination  \citep{chen2015development, ho2019feather}. In animal ecology, space use patterns are governed in part by proximity to resources that are fundamental for survival (e.g. food, water, shelter) \citep{aarts2008estimating, boyce2016can, van2016movement}. Yet many species are highly social and show attraction towards conspecifics, as well as being attracted to familiar areas (a phenomenon called `home ranging', that can be viewed as a form of aggregation \citep{briscoeetal2002, borgeretal2008}). Therefore, like the cellular case, the space use patterns of animals emerge from a combination of self-attraction and attraction to environmental resources \citep{horne2008synoptic,  potts2023scale}.  Although there are some studies incorporating both self-attraction  and environmental effects in the context of PDEs for specific biological situations, for example locust foraging \citep{georgiou2021modelling} and white-tailed deer space use \citep{ellison2024combining}, it would be valuable to build a general theory of how self attraction and heterogeneous landscapes combine to shape the overall space use of organisms.

As a first step towards this end, we study the aggregation-diffusion equation in a static heterogeneous environment in one spatial dimension. It is possible to gain analytic insight into the steady state solutions under certain conditions, made for mathematical tractability. Specifically, we first assume diffusion is quadratic and that the environment can be decomposed as a Fourier series (and is piecewise twice-differentiable). Then we either assume that the non-local advection term has a particular functional form that allows for exact analysis (namely the Laplace kernel) or approximate the system via a Taylor expansion closed at the second moment. These two models are detailed in Section \ref{sec:model}. With these conditions in place, the steady state solutions to the system are fully classified in Section \ref{sec:ssem}. 

Next, to understand which of these steady state solutions are likely to be observed in reality (i.e. in numerical experiments), we examine the energy functional associated to the system and minimise it across the possible steady states. In Section \ref{sec:scl} we perform this minimisation procedure for a particular functional form of the environment, related to a single clump of attractive resources. In the case of cells, this could be thought of as a chemical gradient arising from a point source. For animals, this models an area of high forage in amongst low-forage surroundings. The energy minimisation procedure turns out to be very rapid, as it only involves searching through a single variable across a finite range of values. This enables us to ascertain quickly how the functional form of the minimum energy steady state solution varies with the model parameters, without the need for time-consuming numerical PDEs. 

In Section \ref{sec:num}, we explore numerically the extent to which the lessons from our analytic study extend to situations that are not amenable to mathematical analysis. Specifically, we focus on different functional forms for the non-local kernel and linear diffusion alongside quadratic. Linear diffusion, whilst not so mathematically amenable as quadratic diffusion (at least in our case), is perhaps more natural biologically. So it is interesting to see whether our analytic results carry over to different situations that may be slightly closer to the underlying biological reality. Finally, Section \ref{sec:ni2} gives some preliminary numerical exploration of cases adjacent to those studied here, pointing towards possible future extensions of the present work.

\section{The model}\label{sec:model}

Let $u(x,t)$ denote the population density across space, $x$, of a group of organisms at time $t$. The total population is assumed to be of fixed size, so that any patterns that form are governed purely by the organisms' movement (rather than changes in the overall population size). The organisms each have a diffusive aspect to their movement, as well as non-local attraction to other organisms, and a tendency to move up the gradient of a fixed environment, given by $A(x)$. (Note that, if we are interpreting the environment as resources, the assumption is that any resource decay happens much slower than the timescale over which the population density reaches a steady state.) In one spatial dimension, this leads to the following model
\begin{linenomath*}\begin{align}
		\label{eq:aggdiff_gen}
		\frac{\partial u}{\partial t}&=\frac{D}{k}\frac{\partial^2}{\partial x^2}(u^k)- \frac{\partial}{\partial x}\left[ u\left(\gamma\frac{\partial}{\partial x} (K\ast u)+\frac{\partial A}{\partial x}\right)\right],
\end{align}\end{linenomath*}
where $D,\gamma>0$ are positive real constants, $k>0$ is a constant integer, $K(x)$ is a symmetric probability density function (so integrates to 1 over the domain of definition) with finite variance, and 
\begin{linenomath*}\begin{align}
		\label{eq:conv}
		K\ast u(x) = \int_{-\infty}^{\infty}  K(z)u(x+z){\rm d}z
\end{align}\end{linenomath*}
is a convolution.  Equation (\ref{eq:aggdiff_gen}) is an aggregation-diffusion equation \citep{carrillo2019aggregation}, with an additional term denoting flow up the gradient of the environment, $A(x)$. Notice that, although the organisms' movement with respect to the environment is described in an ostensibly local way in Equation (\ref{eq:aggdiff_gen}), $A(x)$ could represent a nonlocal averaging of an underlying environment \citep{fagan2017perceptual}. For example, if the environment were described by a function ${\mathcal A}(x)$ and the organism's nonlocal perception  of the environment were given by ${\mathcal K}(x)$ then one could set $A(x)={\mathcal K}\ast {\mathcal A}(x)$.

For most of this manuscript, in particular for deriving all the analytic results, we will be focusing on the quadratic diffusion case, $k=2$, which rearranges to the following form
\begin{linenomath*}\begin{align}
		\label{eq:aggdiff}
		\frac{\partial u}{\partial t}&=\frac{\partial}{\partial x}\left[u\left(D\frac{\partial u}{\partial x}-\gamma \frac{\partial}{\partial x} (K\ast u)-\frac{\partial A}{\partial x}\right)\right].
\end{align}\end{linenomath*}
The non-local term, $K\ast u$, in Equation (\ref{eq:aggdiff}) makes analytic studies tricky in general. However, there are two things we can do to ease matters. First, it turns out that the Laplace kernel has some nice properties that allow for exact analysis (namely being the solution to a particular differential operator: see Proposition \ref{thm:ss1} later). We denote the Laplace kernel by $K_m(x)=m{\rm e}^{-m|x|}/2$ to separate it from the general kernel, $K$. The standard deviation of the Laplace distribution is $\sigma=\sqrt{2}/m$, which can be viewed as measuring the extent of the non-local sensing of the organism \citep{fagan2017perceptual}.

Second, for any $K$ that is symmetric and has finite variance, we can make the following second order approximation
\begin{linenomath*}\begin{align}
		\label{eq:approx}
		K\ast u \approx u+\frac{\sigma^2}{2}\frac{\partial^2 u}{\partial x^2},
\end{align}\end{linenomath*}
where $\sigma$ is the standard deviation of $K$ (note that symmetry of $K$ means it has zero mean).  This approximation is derived by taking a Taylor expansion of $u(x,t)$, and assumes that the moments of the distribution $K(x)$ decay sufficiently fast. See \citet[Section 2.1]{falco2023local} for a more detailed derivation.

For our purposes, this approximation enables us to replace Equation (\ref{eq:aggdiff}) with the following local fourth-order equation
\begin{linenomath*}\begin{align}
		\label{eq:adfo}
		\frac{\partial u}{\partial t}&=\frac{\partial}{\partial x}\left[u\left(D\frac{\partial u}{\partial x}-\gamma \frac{\partial u}{\partial x}-\frac{\gamma\sigma^2}{2} \frac{\partial^3u}{\partial x^3}-\frac{\partial A}{\partial x}\right)\right],
\end{align}\end{linenomath*}
which can be viewed as a Cahn-Hilliard equation \citep{kim2016basic} with added environmental heterogeneity. Subsequent analysis will focus on functional forms for $A(x)$ that are periodic on the interval $[-L,L]$ (i.e. $A(x)=A(x+2L)$ for all $x \in {\mathbb R}$) so they can be decomposed as the following Fourier series 
\begin{linenomath*}\begin{align}
		\label{eq:ax}
		A(x)=a_0+\sum_{n=1}^\infty\left[a_n\cos\left(\frac{n\pi x}{L}\right)+b_n\sin\left(\frac{n\pi x}{L}\right)\right].
\end{align}\end{linenomath*}
To ease analysis, we will use the following non-dimensionalisation
\begin{linenomath*}\begin{align}
		\label{eq:nondim}
		\tilde{x}=\frac{x}{L},\,\tilde{\sigma}=\frac{\sigma}{L},\,\tilde{\gamma}=\frac{\gamma}{D},\,\tilde{A}(\tilde{x})=\frac{LA(x)}{D},\,\tilde{t}=\frac{tD}{L^3},\,\tilde{u}=Lu,\,\tilde{a}_n=\frac{La_n}{D},\,\tilde{b}_n=\frac{Lb_n}{D},\,\tilde{m}=mL.
\end{align}\end{linenomath*}	
Immediately dropping the tildes for notational convenience leads to the following dimensionless versions of Equation (\ref{eq:aggdiff}), written here with $K=K_m$
\begin{linenomath*}\begin{align}
		\label{eq:aggdiffnd}
		\frac{\partial u}{\partial t}&=\frac{\partial}{\partial x}\left(u\frac{\partial }{\partial x}\left[u-\gamma  (K_m\ast u)-A \right]\right),
\end{align}\end{linenomath*}
and the following dimensionless version of Equation (\ref{eq:adfo})
\begin{linenomath*}\begin{align}
		\label{eq:adfond}
		\frac{\partial u}{\partial t}&=\frac{\partial}{\partial x}\left(u\frac{\partial}{\partial x}\left[(1-\gamma)u-\frac{\gamma\sigma^2}{2} \frac{\partial^2u}{\partial x^2}-A\right]\right),
\end{align}\end{linenomath*}
where
\begin{linenomath*}\begin{align}
		\label{eq:axnd}
		A(x)=a_0+\sum_{n=1}^\infty\left[a_n\cos\left({n\pi x}\right)+b_n\sin\left({n\pi x}\right)\right].
\end{align}\end{linenomath*}
Equations (\ref{eq:aggdiffnd}) and (\ref{eq:adfond}) will be the main study equations for Sections \ref{sec:ssem} and \ref{sec:scl}.  Note that these equations both preserve total mass across the real line. This mass is defined as
\begin{linenomath*}\begin{align}
		\label{eq:p}
		p=\int_{-\infty}^\infty u(x,t) {\rm d}x.
\end{align}\end{linenomath*}


\section{Steady states and energy minimisers}
\label{sec:ssem}

Here, we classify all the steady state solution to Equations (\ref{eq:aggdiffnd}) and (\ref{eq:adfond}). The results are summarised in two propositions.
\\
\begin{proposition}
	Suppose $u_\ast(x)$ is a steady state solution to Equation (\ref{eq:aggdiffnd}), with $A(x)$ as given in Equation (\ref{eq:axnd}). Suppose $\gamma\neq 1$ and the support of $u_\ast(x)$ is bounded. On any connected component of the support of $u_\ast(x)$, the following holds
    \begin{linenomath*}\begin{align}
		\label{eq:ustarx}
		u_\ast(x)=u_P(x)+u_I(x),
\end{align}\end{linenomath*}
where 
\begin{linenomath*}\begin{align}
		\label{eq:up}
		u_P(x)&=\alpha_0+\sum_{n=1}^\infty\frac{n^2 \pi^2+m^2}{n^2 \pi^2+m^2(1-\gamma)}\left[a_n\cos\left(n\pi x\right)+b_n\sin\left(n\pi x\right)\right], \\
		\label{eq:ui}
u_I(x)&=\begin{cases}P\sin\left(x\sqrt{m^2(\gamma-1)}\right)+Q\cos\left(x\sqrt{m^2(\gamma-1)}\right),&\mbox{if $\gamma>1$},\\
	P\exp\left(x\sqrt{m^2(1-\gamma)}\right)+Q\exp\left(-x\sqrt{m^2(1-\gamma)}\right),&\mbox{if $\gamma<1$},\end{cases}
\end{align}\end{linenomath*}
and $P,Q,\alpha_0 \in {\mathbb R}$ are arbitrary constants.
\label{thm:ss1}
\end{proposition}
\noindent{\bf Proof. }Steady states of Equation (\ref{eq:aggdiffnd}), denoted by $u_\ast(x)$, satisfy 
\begin{linenomath*}\begin{align}
		\label{eq:aggdiff_ssl}
		C&=u_\ast\frac{{\rm d}}{{\rm d} x}\left[u_\ast-\gamma(K_m\ast u_\ast)-A\right],
\end{align}\end{linenomath*}
for some constant $C$.  As the support of $u_\ast(x)$ is bounded, the flux is zero sufficiently far from the origin, so $C=0$.  Hence
\begin{linenomath*}\begin{align}
		\label{eq:aggdiff_ssl2}
		0&=u_\ast\frac{{\rm d} }{{\rm d} x}\left[u_\ast-\gamma (K_m\ast u_\ast)-A\right].
\end{align}\end{linenomath*}
Then, on any connected component of the support of $u_\ast(x)$ (i.e. where $u_\ast(x) \neq 0$), we have 
\begin{linenomath*}\begin{align} 
		\label{eq:aggdiff_sslz2}
		c+A&=u_\ast-\gamma (K_m\ast u_\ast),
\end{align}\end{linenomath*}
for some constant $c$.  Now we apply a particular property of the Laplace kernel, namely that it satisfies
\begin{linenomath*}\begin{align}
		\frac{{\rm d}^2 K_m}{{\rm d} x^2}-m^2K_m+m^2\delta(x)=0,
		\label{eq:lap_loc}
\end{align}\end{linenomath*}
where $\delta(x)$ is the Dirac delta function.  Applying the operator $m^2-\frac{{\rm d}^2}{{\rm d} x^2}$ to Equation (\ref{eq:aggdiff_sslz2}) gives
\begin{linenomath*}\begin{align}
		\label{eq:aggdiff_ss}
		m^2c+m^2A(x)-\frac{{\rm d}^2 A}{{\rm d} x^2}&=m^2(1-\gamma)u_*-\frac{{\rm d}^2 u_*}{{\rm d} x^2},
\end{align}\end{linenomath*}
which is an inhomogeneous second order ODE with constant coefficients. A direct calculation shows that Equation (\ref{eq:aggdiff_ss}) is solved by Equations (\ref{eq:ustarx})-(\ref{eq:ui}) by setting $\alpha_0=(c+a_0)/(1-\gamma)$.
\qed
\\
\begin{proposition}
	Suppose $u_\ast(x)$ is a steady state solution to Equation (\ref{eq:adfond}), with $A(x)$ as given in Equation (\ref{eq:axnd}).  Suppose $\gamma\neq 1$ and the support of $u_\ast(x)$ is bounded. On any connected component of the support of $u_\ast(x)$, the following holds
	\begin{linenomath*}\begin{align}
			\label{eq:ustarx4}
			u_\ast(x)=u_P(x)+u_I(x),
	\end{align}\end{linenomath*}
	where 
	\begin{linenomath*}\begin{align}
			\label{eq:up4}
			u_P(x)&=\alpha_0+\sum_{n=1}^\infty\frac{2}{2(1-\gamma)+\gamma\sigma^2n^2\pi^2}\left[\alpha_n\cos\left(n\pi x\right)+\beta_n\sin\left(n\pi x\right)\right], \\
			\label{eq:ui4}
			u_I(x)&=\begin{cases}P\sin\left(x\sqrt{\frac{2(\gamma-1)}{\gamma\sigma^2}}\right)+Q\cos\left(x\sqrt{\frac{2(\gamma-1)}{\gamma\sigma^2}}\right),&\mbox{if $\gamma>1$},\\
				P\exp\left(x\sqrt{\frac{2(1-\gamma)}{\gamma\sigma^2}}\right)+Q\exp\left(-x\sqrt{\frac{2(1-\gamma)}{\gamma\sigma^2}}\right),&\mbox{if $\gamma<1$},\end{cases}
	\end{align}\end{linenomath*}
	and $P,Q,\alpha_0 \in {\mathbb R}$ are arbitrary constants.
\label{thm:ss2}
\end{proposition}
\noindent{\bf Proof. } Any steady state, $u(x,t)=u_\ast(x)$, of Equation (\ref{eq:adfond}) satisfies
\begin{linenomath*}\begin{align}
		\label{eq:adfo_ss}
		C&=u_\ast\frac{{\rm d} }{{\rm d} x}\left[(1-\gamma)u_\ast-\frac{\gamma\sigma^2}{2} \frac{{\rm d}^2u_\ast}{{\rm d} x^2}-A\right],
\end{align}\end{linenomath*}
for some constant $C$.  Since the support of $u_\ast(x)$ is bounded, the flux is zero sufficiently far from the origin, so $C=0$.  Therefore, as in the proof of Proposition \ref{thm:ss1}, on any connected component of the support of $u_\ast$, we have 
\begin{linenomath*}\begin{align}
		\label{eq:adfo_ssz2}
		c+A=(1-\gamma) u_\ast-\frac{\gamma\sigma^2}{2} \frac{{\rm d}^2 u_\ast}{{\rm d} x^2},
\end{align}\end{linenomath*}
for some constant $c$. A direct calculation shows that Equation (\ref{eq:adfo_ssz2}) is solved by Equations (\ref{eq:ustarx4})-(\ref{eq:ui4}) by setting $\alpha_0=(c-a_0)/(\gamma-1)$.
\qed
\\
\\
{\bf Remark 1.} It is also possible to find a solution to the above propositions in the singular case $\gamma=1$ (Appendix \ref{sec:AppC}).
\\
\\
{\bf Remark 2.} Equations (\ref{eq:ui}) and (\ref{eq:ui4}) arise from the aggregation term, whereas Equations (\ref{eq:up}) and (\ref{eq:up4}) come from the environmental heterogeneity. So they can be viewed as the contributions of self-aggregation and organism-environment interaction, respectively,  to the steady state solution.  
\\
\\ 
Whilst the two above propositions fully-categorise all possible steady state solutions, $u_\ast(x)$, the story is not finished. First, there are three unknowns that the results introduce: $\alpha_0$, $P$, $Q$. Furthermore, the expressions in Equations (\ref{eq:ustarx})-(\ref{eq:ui}) and (\ref{eq:ustarx4})-(\ref{eq:ui4}) are only valid on connected components of the support of $u_\ast(x)$. This leaves open the question as to which of the various possible steady states the PDE system might actually tend towards, given an initial condition. 

To gain insight into this, we look for solutions that minimise the associated energy functional \citep{giunta2022detecting}. For Equation (\ref{eq:aggdiffnd}), this functional is
\begin{linenomath*}\begin{align}
		\label{eq:energy}
		E_1[u]=\int_{-\infty}^\infty u\left[(1-\gamma)u-2A-\gamma K_m*u\right]{\rm d}x.
\end{align}\end{linenomath*}	
This is a slight modification of analogous energy functional constructed in the homogeneous case, e.g. \citet[Equation 5]{carrillo2019aggregation}.  A direct calculation shows that 
\begin{linenomath*}\begin{align}
		\label{eq:energy_change}
		\frac{{\rm d}E_1}{{\rm d}t}=-\int_{-\infty}^\infty 2u\left(\frac{\partial}{\partial x}\left[u-A-\gamma K_m*u\right]\right)^2{\rm d}x,
\end{align}\end{linenomath*}	
as long as $u(x)$ vanishes for all $x$ arbitrarily far from the origin. This means that $E_1[u]$ is non-increasing in time as long as $u$ remains non-negative, and is zero when Equation (\ref{eq:aggdiffnd}) is at steady state.

The energy for Equation (\ref{eq:adfond}) is
\begin{linenomath*}\begin{align}
		\label{eq:energy2}
		E_2[u]=\int_{-\infty}^\infty u\left[(1-\gamma)u-2A-\frac{\gamma\sigma^2}{2}\frac{\partial^2u}{\partial x^2}\right]{\rm d}x.
\end{align}\end{linenomath*}	
Similarly, a direct calculation shows that 
\begin{linenomath*}\begin{align}
		\label{eq:energy_change2}
		\frac{{\rm d}E_2}{{\rm d}t}=-\int_{-\infty}^\infty 2u\left(\frac{\partial}{\partial x}\left[(1-\gamma)u-A-\frac{\gamma\sigma^2}{2}\frac{\partial^2u}{\partial x^2}\right]\right)^2{\rm d}x,
\end{align}\end{linenomath*}	
as long as $u(x)$ vanishes for all $x$ arbitrarily far from the origin.
Therefore, as before, $E_2[u]$ is non-increasing in time as long as $u$ remains non-negative, and is zero when Equation (\ref{eq:adfond}) is at steady state.

\begin{figure}[h!]
	\begin{center}
		\includegraphics[width=\columnwidth]{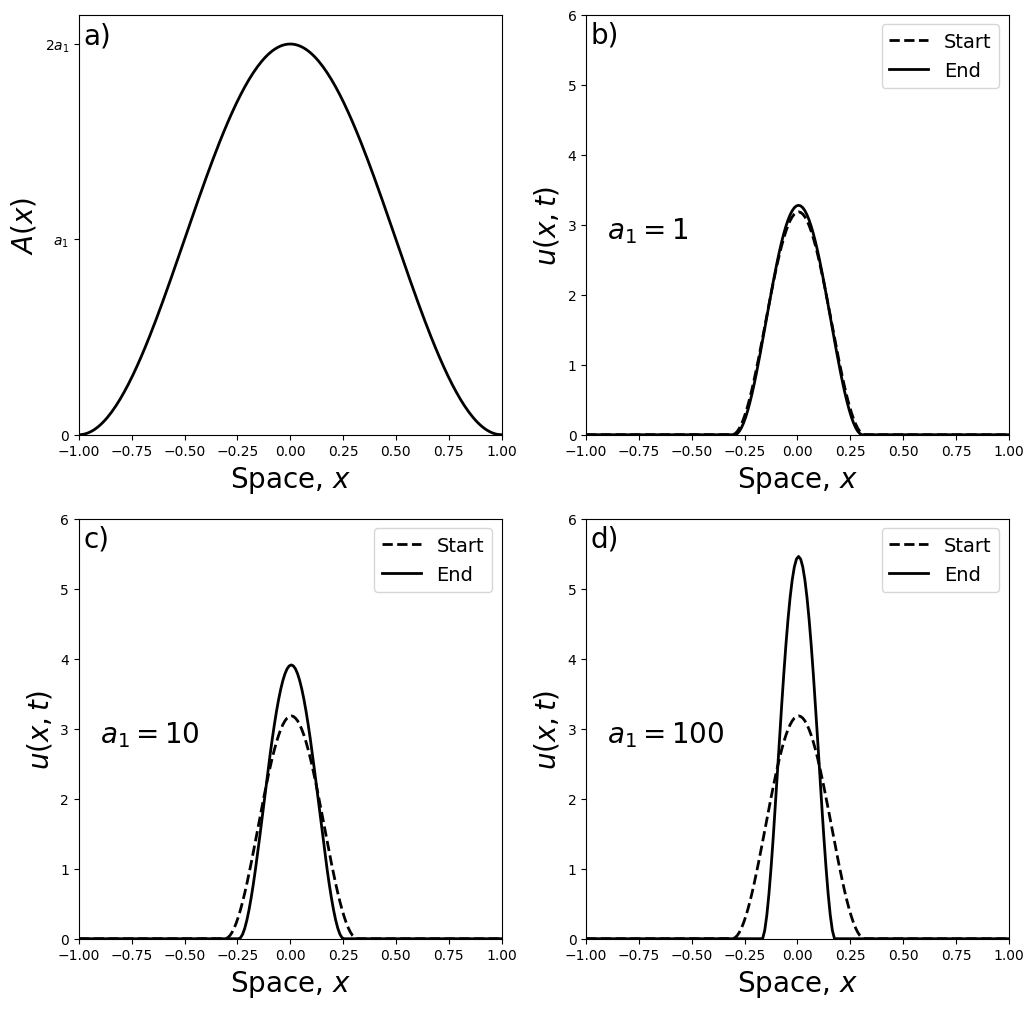}
	\end{center}
	\caption{{\bf Aggregation in a single-clumped landscape: numerics.}  Panel (a) shows $A(x)=a_1[1+\cos(\pi x)]$, a single clump of attractive resources centred on $x=0$. With this functional form of $A(x)$ in place, Panels (b)-(d) show initial (`Start') and final (`End') numerical solutions for Equation (\ref{eq:adfo}), for example values of $a_1$. The initial condition is the minimum energy solution in the case where $A(x)=0$ (a homogeneous landscape), so we can see how the introduction of landscape heterogeneity affects the shape of the aggregation.}
	\label{fig:adh_start_end_plots4}
\end{figure}

\section{Landscapes with a single clump of attractive resources}
\label{sec:scl}

As mentioned in the Introduction, we focus our attention on a particular case of biological interest: where the landscape $A(x)$ consists of a single clump of attractive material. 
The aim is to disentangle the landscape and self-attractive effects on the size and shape of the resulting aggregation. The functional form for $A(x)$ we use to model this situation is as follows
\begin{linenomath*}\begin{align}
		\label{eq:ax_1hump}
		A(x)=\begin{cases}a_n[1+\cos(n\pi x)] , & \mbox{if $|x|<\frac{1}{n}$,}\\
			0, &\mbox{otherwise.}\end{cases}
\end{align}\end{linenomath*}
Our analysis will focus on the fourth-order model of Equation (\ref{eq:adfond}). Calculations for the Laplace model of Equation (\ref{eq:aggdiffnd}) are rather similar, so we report these in Appendix \ref{sec:AppA}. 

We will also focus on the case $\gamma>1$, as (i) this is above the Turing threshold for spontaneous pattern formation in Equation (\ref{eq:adfond}) with no resources (i.e. $A(x)=0$), and (ii) this is where the aggregation size is an interesting combination of the sinusoidal terms given by the aggregative tendencies (Equation \ref{eq:ui4}) and the resources (Equation \ref{eq:up4}), so we see non-trivial results. 

In numerical experiments, if we start with aggregated initial conditions, we remain in an aggregation, albeit one of a different size. This is shown in Figure \ref{fig:adh_start_end_plots4}. In this figure, the simulations use an initial condition that is equal to the steady-state minimum-energy solution of Equation (\ref{eq:adfond}) in the case of no resources (i.e. $A(x)=0$) and also $\gamma=2$, $\sigma =0.1$, and total population size $p=1$. This function has the following form [first shown in \citet[Section 2.3.1]{falco2023local}, but also reported in Appendix \ref{sec:AppB} with a slightly different proof]
\begin{linenomath*}\begin{align}
	\label{eq:ustarx_homr0}
	u(x,0)=u_\ast(x)|_{A(x)=0}=\begin{cases}\frac{p}{2r_0}\left[1+\cos\left(\frac{x\pi}{r_0}\right)\right], & \mbox{if $-r_0<x<r_0$,}\\
			0, &\mbox{otherwise,}\end{cases}
\end{align}\end{linenomath*}
where
\begin{linenomath*}\begin{align}
	\label{eq:r0}
	r_0 = \pi\sqrt{\frac{\gamma\sigma^2}{2(\gamma-1)}},
\end{align}\end{linenomath*}
and $p$ is as defined in Equation (\ref{eq:p}).

From the initial condition given by Equation (\ref{eq:ustarx_homr0}), Figure \ref{fig:adh_start_end_plots4} shows numerical steady-state  solutions of Equation (\ref{eq:adfond}), calculated by numerically solving this PDE through time using the algorithm from \citet{falco2023local}, keeping $\gamma=2$ and $\sigma=0.2$ fixed but using different values of $a_1$. Each steady state was estimated by running the algorithm until $|u(x,t)-u(x,t+\Delta t)|<10^{-8}$ for all $x$, where $\Delta t=10^{-7}$.

	\begin{figure}[h!]
	\begin{center}
		\includegraphics[width=\columnwidth]{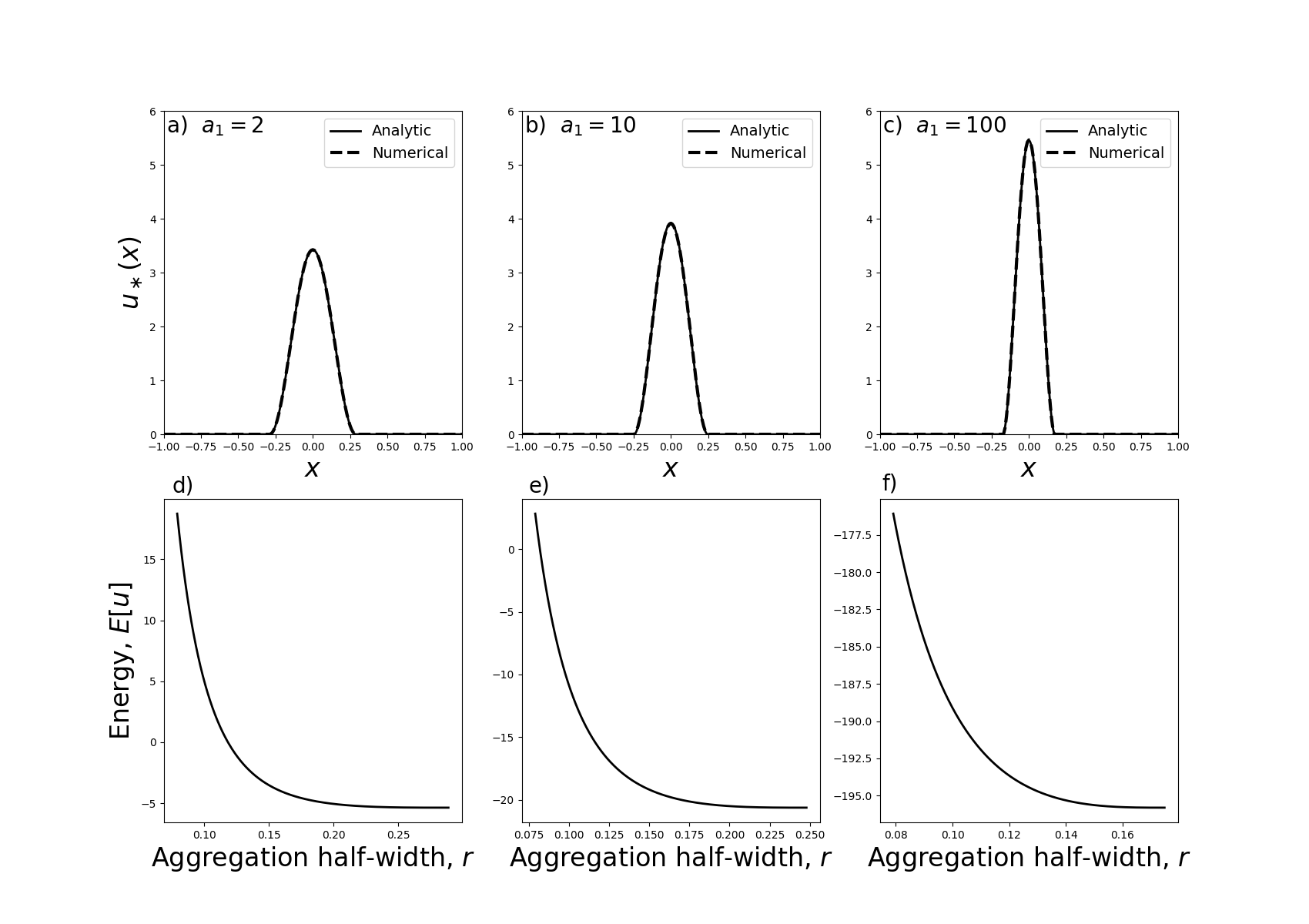}
	\end{center}
	\caption{{\bf Aggregation in a single-clump landscape: minimum energy solutions.}  Panels (a), (b), and (c) show the minimum energy solution of the form given by Equations (\ref{eq:ustarx_1hump})-(\ref{eq:1hump_Qcts}) (solid curves), for $n=1$, alongside the numerical steady-state solution (dashed curves) for different values of $a_1$ given in the plots.  Panels (d), (e), and (f) show the energy as a function of $r$ for the values of $a_1$ given in Panels (a), (b), and (c) respectively.  The respective minimum energy $r$-values are $r=0.289, r=0.248,$ and $r=0.175$.}
	\label{fig:adh_plot_ss_energy}
\end{figure} 
Each numerical steady-state solution appears to be supported on a single interval $[-r,r]$, symmetric about zero.  We therefore restrict our search for minimum energy solutions, from the possible solutions found in Section \ref{sec:ssem}, to this type of symmetric, single-aggregation solution. In the fourth-order model of Equation (\ref{eq:adfond}), such solutions have the following form (derived from Equations (\ref{eq:ustarx4})-(\ref{eq:ui4}) and (\ref{eq:ax_1hump})),
    \begin{linenomath*}\begin{align}
		\label{eq:ustarx_1hump}
		u_\ast(x)=\begin{cases}\alpha_0+\alpha_n[1+\cos(\pi nx)]+Q\cos\left(\frac{x\pi}{r_0}\right), & \mbox{for $x \in \left[-\frac{1}{n},\frac{1}{n}\right]\cap[-r,r]$,}\\
			\alpha_0+Q\cos\left(\frac{x\pi}{r_0}\right), & \mbox{for $\frac{1}{n}<|x|\leq r$ if $r>\frac{1}{n}$,}\\
			0, &\mbox{otherwise,}\end{cases}
\end{align}\end{linenomath*}
where $0<r\leq 1$. Applying the integral condition from Equation (\ref{eq:p}) gives
\begin{linenomath*}\begin{align}
		\label{eq:1hump_alpha0}
		\alpha_0=\begin{cases}\frac{p}{2r}-\alpha_n - \frac{\alpha_n}{\pi n r}\sin(\pi n r)-\frac{Qr_0}{\pi r}\sin\left(\frac{r\pi}{r_0}\right),&\mbox{if $r<1/n$,} \\
			\frac{p}{2r}-\frac{\alpha_n}{n r}-\frac{Qr_0}{\pi r}\sin\left(\frac{r\pi}{r_0}\right),&\mbox{otherwise,}
		\end{cases}
\end{align}\end{linenomath*}
so that the two remaining free parameters are $r$ and $Q$. We restrict our search further by only looking for continuous solutions. 
\begin{figure}[h!]
	\begin{center}
		\includegraphics[width=\columnwidth]{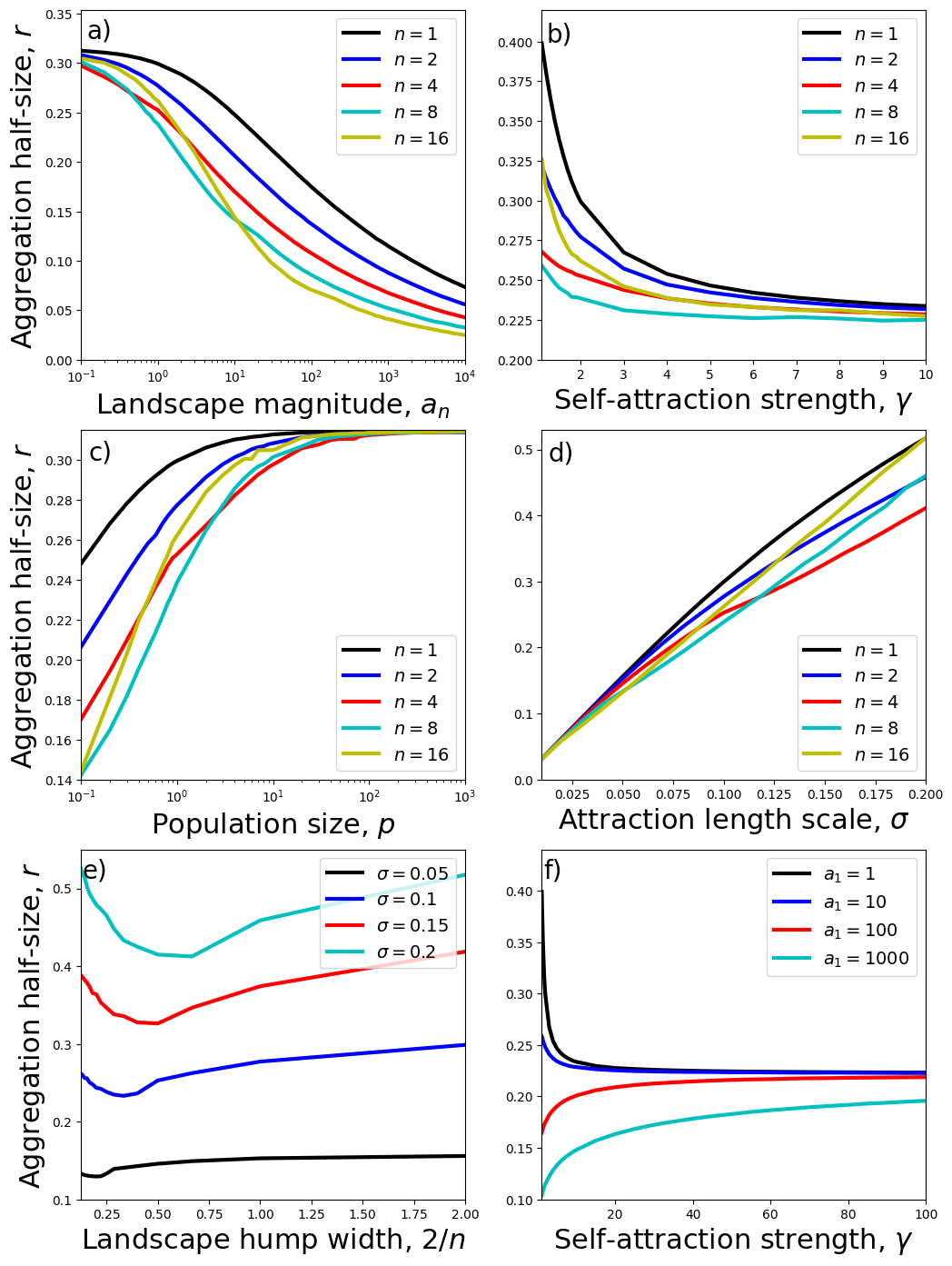}
	\end{center}
	\caption{{\bf Dependence of aggregation size on model parameters.} Aggregation sizes refer to minimum energy steady-state solutions to Equation (\ref{eq:adfond}) with $A(x)$ as given in Equation (\ref{eq:ax_1hump}), and calculated using the energy-minimisation procedure from Section \ref{sec:scl}. Unless otherwise stated, $n=1$, $a_n=1$, $\gamma=2$, and $\sigma=0.1$. }
	\label{fig:adh_plot_r_vs_params}
\end{figure} 
This gives the additional constraint
\begin{linenomath*}\begin{align}
		\label{eq:1hump_Qcts}
		Q = \begin{cases}-\frac{\alpha_0+\alpha_n+\alpha_n\cos(\pi n r)}{\cos(r\pi/r_0)}, & \mbox{if $r<1/n$,} \\
			-\frac{\alpha_0}{\cos(r\pi/r_0)},&\mbox{otherwise.}
		\end{cases}
\end{align}\end{linenomath*}
Plugging the expression for $u=u_\ast$ from Equation (\ref{eq:ustarx_1hump}) into Equation (\ref{eq:energy2}) shows, after a direct calculation, that the energy functional we wish to minimise has the form
\begin{linenomath*}\begin{align}
		\label{eq:1hump_energy}
		E_2[u_\ast] = \begin{cases}\int_{-r}^{r}u_\ast[(1-\gamma)(\alpha_0+\alpha_n)-2a_n-a_n\cos(n\pi x)]{\rm d}x, & \mbox{if $r<1/n$,} \\
			\int_{-1/n}^{1/n} u_\ast[(1-\gamma)(\alpha_0+\alpha_n)-2a_n-a_n\cos(n\pi x)]{\rm d}x \\
			\qquad+2\int_{1/n}^r u_\ast(1-\gamma)\alpha_0{\rm d}x,&\mbox{otherwise.}
		\end{cases}
\end{align}\end{linenomath*}
Minimising Equation (\ref{eq:1hump_energy}) requires a numerical search through the single remaining parameter, $r$. This can be done in a fraction of a second on an ordinary laptop (e.g. one with an Intel i7 2.8GHz processor), contrasting with numerical solutions to the underlying PDE, which typically take many minutes or even hours. Furthermore, once $r$ is found numerically, the minimum energy solution can be written down in an analytic form, namely that of Equations (\ref{eq:ustarx_1hump}).

Figure \ref{fig:adh_plot_ss_energy} shows a few examples of such analytic solutions, together with their numerical counterparts, solved using the finite-volume algorithm of \citet{bailo2023unconditional} and \citet{falco2023local}. Note that for certain values of $r\in(0,r_0)$, the solution given by Equations (\ref{eq:ustarx_1hump})-(\ref{eq:1hump_Qcts}) is not positive, so not  allowable (see the positivity results of \citet{bailo2020fully}, \citet{bailo2023unconditional}). Therefore the horizontal axes in Figures \ref{fig:adh_plot_ss_energy}(d-f) do not go all the way from 0 to $r_0$.  

\begin{figure}[h!]
	\begin{center}
		\includegraphics[width=\columnwidth]{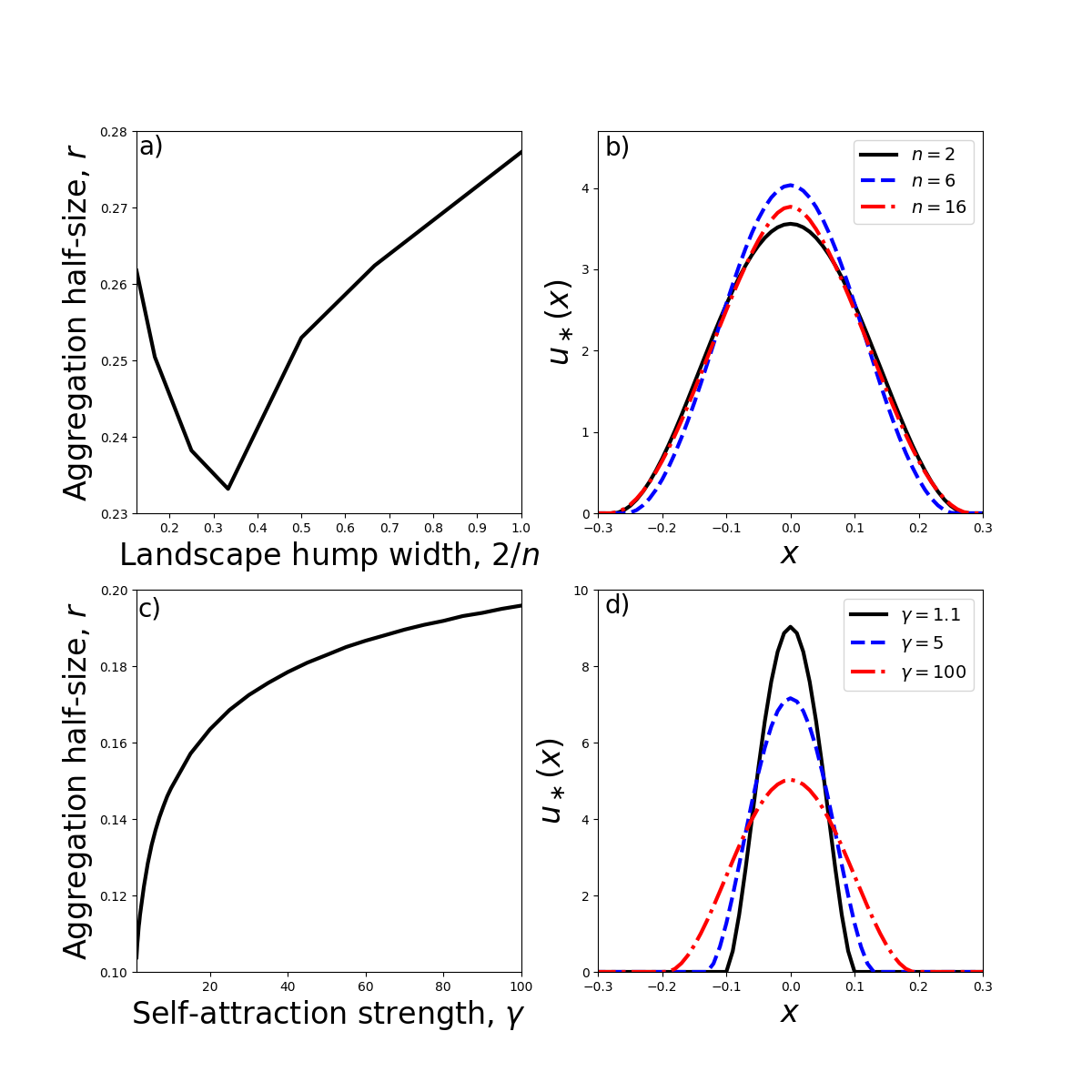}
	\end{center}
	\caption{{\bf Numerical verification of analytic insights. } Panel (a) shows a zoomed-in version of the case $\sigma=0.1$ from Figure \ref{fig:adh_plot_r_vs_params}e. Panel (b) shows numerical steady state-solutions to Equation (\ref{eq:adfond}) corresponding to the parameters from Panel (a) (namely, $\gamma=2$, $a_n=1$, $\sigma=0.1$). Panel (c) shows the case $a_1=1000$ from Figure \ref{fig:adh_plot_r_vs_params}f and Panel (d) gives the corresponding numerical steady-state solutions.  }
	\label{fig:adh_plot_ss_n_gamma}
\end{figure} 

This energy minimisation procedure enables rapid calculating of trends in the size of the aggregation as a function of the underlying parameters, without needing to perform time-consuming numerical PDEs.   Figure \ref{fig:adh_plot_r_vs_params} shows how the width of the minimum-energy aggregation of $u$, given by $2r$, depends upon both the width of the landscape's peak, given by $2/n$, as well as the parameters $a_n$, $p$, $\gamma$, and $\sigma$.  
 
Interestingly, there is a non-monotonic dependence of the aggregation width of $u$ on width of the underlying landscape's peak (Figure \ref{fig:adh_plot_r_vs_params}e).  This arises from analysing minimum energy solutions, but is also verified through numerical simulations. In Figure \ref{fig:adh_plot_ss_n_gamma}b, we see that the narrowest aggregation (and also highest peak) occurs for the intermediate value $n=6$. If resources are more clumped than this, i.e. if $n>6$, then rather than this causing the aggregations to be thinner, as might be expected, they are actually slightly wider.

Another counter-intuitive result occurs when we have a strong attraction to resources, e.g. $a_n=100$ or $a_n=1000$. Here, increasing $\gamma$ leads to a widening of the aggregation, contrary to what usually happens with no resources or a smaller resource attraction (Figure \ref{fig:adh_plot_r_vs_params}f). Again, this result plays out in the numerics (Figure \ref{fig:adh_plot_ss_n_gamma}d). Although not shown here, a similar phenomenon also occurs for $n=2,4,8,12,16$ (so narrower resource clumps), each case leading to $r$ being an increasing function of $\gamma$ for $a_n=100$ or $a_n=1000$.

A further interesting feature of Figure \ref{fig:adh_plot_r_vs_params}f is the saturation in the effect of $\gamma$ on aggregation size as $\gamma$ is increased (also observed in Figure \ref{fig:adh_plot_r_vs_params}b). To see why this happens, take the limit as $\gamma \rightarrow \infty$ of the steady state (Equation \ref{eq:adfo_ss}), and notice that the result is independent of the landscape, $A(x)$. Figure \ref{fig:adh_plot_r_vs_params}c also shows a saturation in aggregation size, this time as the population size is increased. In this case, the large population causes self-attraction to dominate over landscape attraction, again rendering the effect of the landscape negligible.

\section{Numerical investigation I: relaxing the core assumptions}
\label{sec:num}

All the analytic results in Sections \ref{sec:ssem} and \ref{sec:scl} rely on two features chosen purely for mathematical convenience: quadratic diffusion ($m=2$ in Equation \ref{eq:aggdiff_gen}) and either a fourth-order approximation to the non-local term or a Laplace kernel. These choices enable us to derive linear ODEs for the steady state that can be solved exactly. However, in many biological applications, linear diffusion is more natural than quadratic diffusion. For example, linear diffusion appears as the continuum limit of many random walk models used for organism movement \citep{hillenpainter2013, patlak1953random, pottslewis2016b, potts2020parametrizing, turchin1989}. Likewise, the Laplace kernel may not always be the most favourable choice from a biological modelling perspective \citep{painter2024biological}. Therefore it is worth investigating whether the analytic insights provided so far might carry over to other situations, including linear diffusion and different non-local kernels.
\begin{figure}[h!]
	\begin{center}
		\includegraphics[width=\columnwidth]{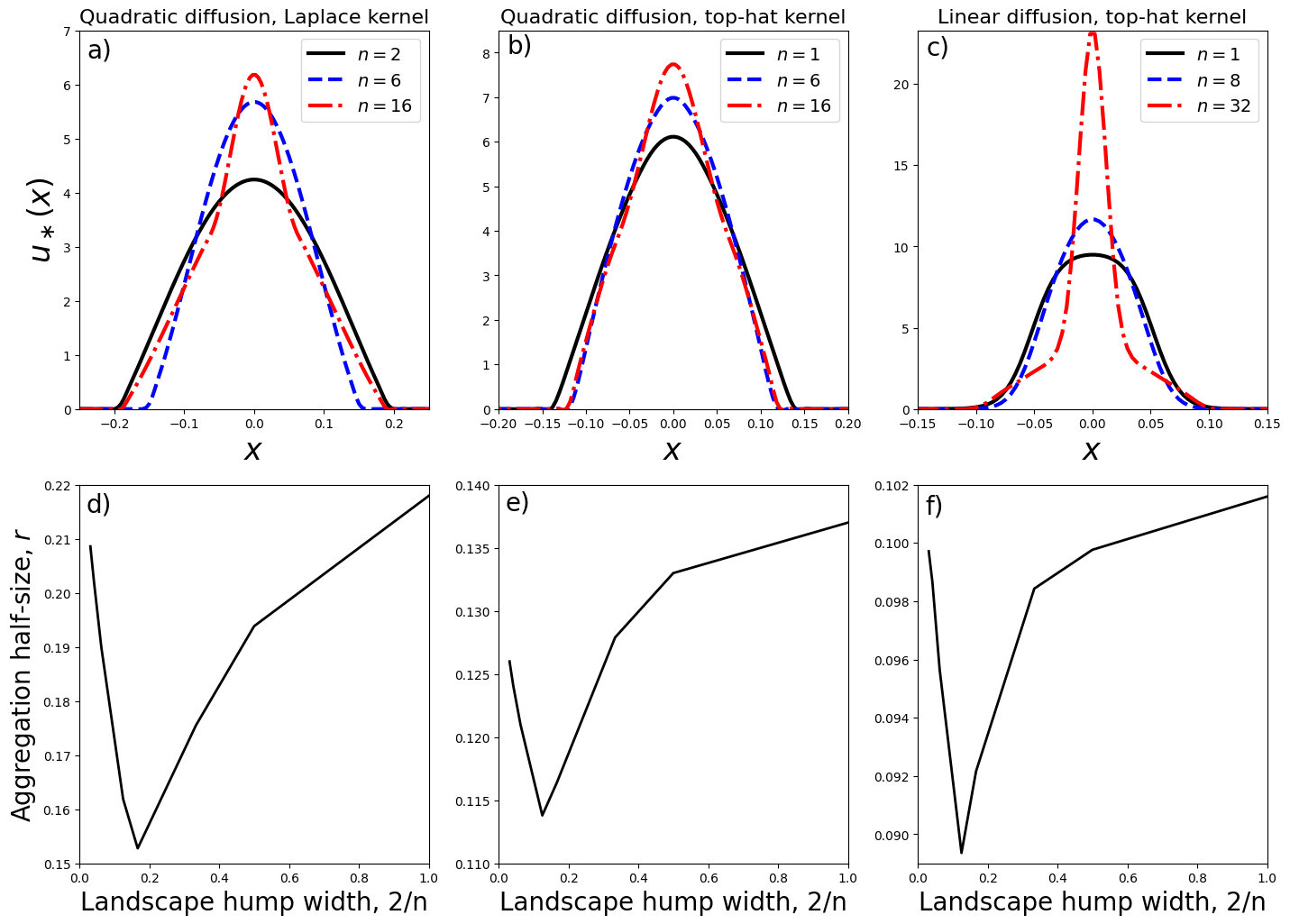}
	\end{center}
	\caption{{\bf Numerical investigation. } Panel (a) shows numerical steady state solutions of Equation (\ref{eq:aggdiffnd}) with a Laplace kernel (Equation \ref{eq:lap_loc}), for various $n$ and $m=10$. Panel (b) shows numerical steady state solutions of Equation (\ref{eq:aggdiffnd}) with $K_m$ replaced with a top hat kernel (Equation \ref{eq:th}), for various $n$ and $\delta=0.1$.  Panel (c) shows numerical steady state solutions of Equation (\ref{eq:aggdiff_lin}) with a top hat kernel with $\delta=0.1$. Panels (d-f) show the aggregation half-size as a function of the resource hump width, corresponding to the numerics shown in Panels (a-c) respectively. In Panels (d-e), the aggregation sizes are calculated from the support of the numerical steady state. For Panel (f), with linear diffusion, we have $u_\ast(x)>0$ across the whole interval $[-1,1]$, so the aggregation sizes are calculated at height $u_\ast(x)=0.1$. In all panels, $\gamma=2$, $a_n=1$, $p=1$, and $A(x)$ is as in Equation (\ref{eq:ax_1hump}).}
	\label{fig:adh_plot_num_ss}
\end{figure} 

The linear diffusion model is given by $k=1$ in Equation (\ref{eq:aggdiff_gen}). After applying the non-dimensionalisation from Equation (\ref{eq:nondim}) and dropping the tildes, this becomes
\begin{linenomath*}\begin{align}
		\label{eq:aggdiff_lin}
		\frac{\partial u}{\partial t}&=\frac{\partial^2 u}{\partial x^2}- \frac{\partial}{\partial x}\left[ u\left(\gamma\frac{\partial}{\partial x} (K\ast u)+\frac{\partial A}{\partial x}\right)\right].
\end{align}\end{linenomath*}
To understand why numerical analysis is required in this case, we can follow the same argument as we did for quadratic diffusion to examine the steady state solutions of Equation (\ref{eq:aggdiff_lin}). Any steady state, $u_\ast(x)$, satisfies the following
\begin{linenomath*}\begin{align}
		\label{eq:aggdiff_ss_lin}
		0&=\frac{{\rm d} }{{\rm d} x}\left[u_\ast\frac{{\rm d} }{{\rm d} x}\left(\ln(u_\ast)-\gamma K \ast u_\ast-A\right)\right].
\end{align}\end{linenomath*}
Then, using a similar argument to that given from Equations (\ref{eq:aggdiff_ssl}) to (\ref{eq:aggdiff_sslz2}), we arrive at the following expression, valid on any connected component of the support of $u_\ast(x)$
\begin{linenomath*}\begin{align}
		\label{eq:aggdiff_ss_lin2}
		c+A&=\ln(u_\ast)-\gamma K \ast u_\ast.
\end{align}\end{linenomath*}
The $\ln(u_\ast)$ term makes this non-linear, removing the possibility of exact mathematical analysis. Indeed, even if we use a fourth order expansion (or a Laplace kernel) to deal with the non-local term, $K \ast u_\ast$, exact solutions are not generally possible.

However, we can use the results from the quadratic diffusion case to guide numerical analysis of the linear diffusion model. In particular, it is interesting to examine whether the non-monotonic dependence of aggregation width, $2r$, on resource-clump width, $2/n$, holds in the case of linear diffusion. Numerical steady-state solutions were examined for both the linear diffusion case (Equation \ref{eq:aggdiff_ss_lin}) and quadratic diffusion (Equation \ref{eq:aggdiffnd}) for two different kernels cases: where $K=K_m$, the Laplace distribution, and where $K_m$ is replaced by $K={\mathcal K}_\delta$, the top-hat distribution given by
\begin{linenomath*}\begin{align}
		\label{eq:th}
		{\mathcal K}_\delta(x)=\begin{cases}\frac{1}{2\delta}, &\mbox{if $|x|<\delta$} \\
			0, &\mbox{otherwise.}
		\end{cases}
\end{align}\end{linenomath*}
The top-hat distribution is chosen due to its popularity in biological modelling \citep{painter2024biological, wang2023open}.  

Results are shown in Figure \ref{fig:adh_plot_num_ss} for three of the four cases. The linear diffusion case with a Laplace kernel is omitted, since we did not see any evidence of the non-monotonic dependence of the resource clump width on the aggregation width in this case. However, we do see this phenomenon in the other three cases. When there is quadratic diffusion and a Laplace kernel (Figure \ref{fig:adh_plot_num_ss}a), the $n=2$ and $n=16$ cases have very similar widths towards the bottom of the aggregation, with $n=6$ markedly thinner. However, contrary to the steady states of the fourth-order local PDE shown in Figure \ref{fig:adh_plot_ss_n_gamma}b, the height of the $n=16$ case is higher than the $n=6$ case. Indeed, we see a notable thinning of the $n=16$ aggregation about about $u_\ast(x)=2.7$. A possible interpretation of this is that towards the bottom of the aggregation, the non-local self-attraction is dominating to push the aggregation towards the width it would be were there no resources, i.e. Equation (\ref{eq:ustarx_homr0}).  Yet in the very centre of the aggregation, the thin resource clump dominates, and we see a change in shape. 

This phenomenon is also present in Figure \ref{fig:adh_plot_num_ss}b, and is even more pronounced in Figure \ref{fig:adh_plot_num_ss}c, where the highest value of $n$ shown is $n=32$. In this case, whilst the aggregation is thicker than the $n=1$ or $n=8$ cases at the bottom, it is quite a bit thinner towards the top, and also much higher. Although we do not have analytic expressions for the plots in Figure \ref{fig:adh_plot_num_ss}, they do look reminiscent of the functional form in Equation (\ref{eq:ustarx_1hump}): a weighted sum of two cosine functions with different widths (truncated when they reach their first minima before and after $0$).

It is also valuable to ask whether there are situations, beyond the fourth-order quadratic-diffusion model (Equation \ref{eq:aggdiffnd}), where we see an increase in aggregation width as the self-attraction increases, like in Figure \ref{fig:adh_plot_ss_n_gamma}c-d, for $a_1=100$ and $a_1=1000$. However, when exploring analogous parameter values to those cases, this counter-intuitive result neither appeared for linear diffusion nor for Laplace or top-hat kernels. This therefore seems to be a particular feature of the fourth-order approximation.

Numerical solutions of the linear diffusion case (Equation \ref{eq:aggdiff_lin}) were computed using a forward difference algorithm by discretising space into a lattice with spacing $\Delta x=0.01$ and time into intervals of length $\Delta t=10^{-5}$. Numerical solutions of the quadratic diffusion PDEs used the algorithm of \citet{bailo2023unconditional} with  $\Delta x=0.01$ and $\Delta t=5\times 10^{-6}$.   To estimate the steady state, each simulation was run until $|u(x,t)-u(x,t+\Delta t)|<10^{-8}$ for all $x$. Code for performing numerics is available on GitHub at \url{https://github.com/jonathan-potts/AggDiffHet}.

\section{Numerical investigation II: different initial conditions and landscapes}
\label{sec:ni2}

Whilst the main purpose of this work is to introduce an analytic technique for ascertaining minimum-energy steady state solutions to Equation (\ref{eq:aggdiffnd}) and (\ref{eq:adfond}), with particular focus on situations where there is a single aggregative steady state, it is valuable to explore numerical solutions away from this example, to showcase some of the other possible patterns that might form. An extensive exploration would be beyond the scope of this paper, but this section gives a few examples to point the way to future numerical studies. The purpose of this section is (i) to highlight the possibility of other local energy minima than those explored analytically in Section \ref{sec:scl}, which may be obtainable through different initial conditions, and (ii) to show how multiple resource clumps might interact with existing aggregations to stretch or break them.
\begin{figure}[h!]
	\begin{center}
		\includegraphics[width=\columnwidth]{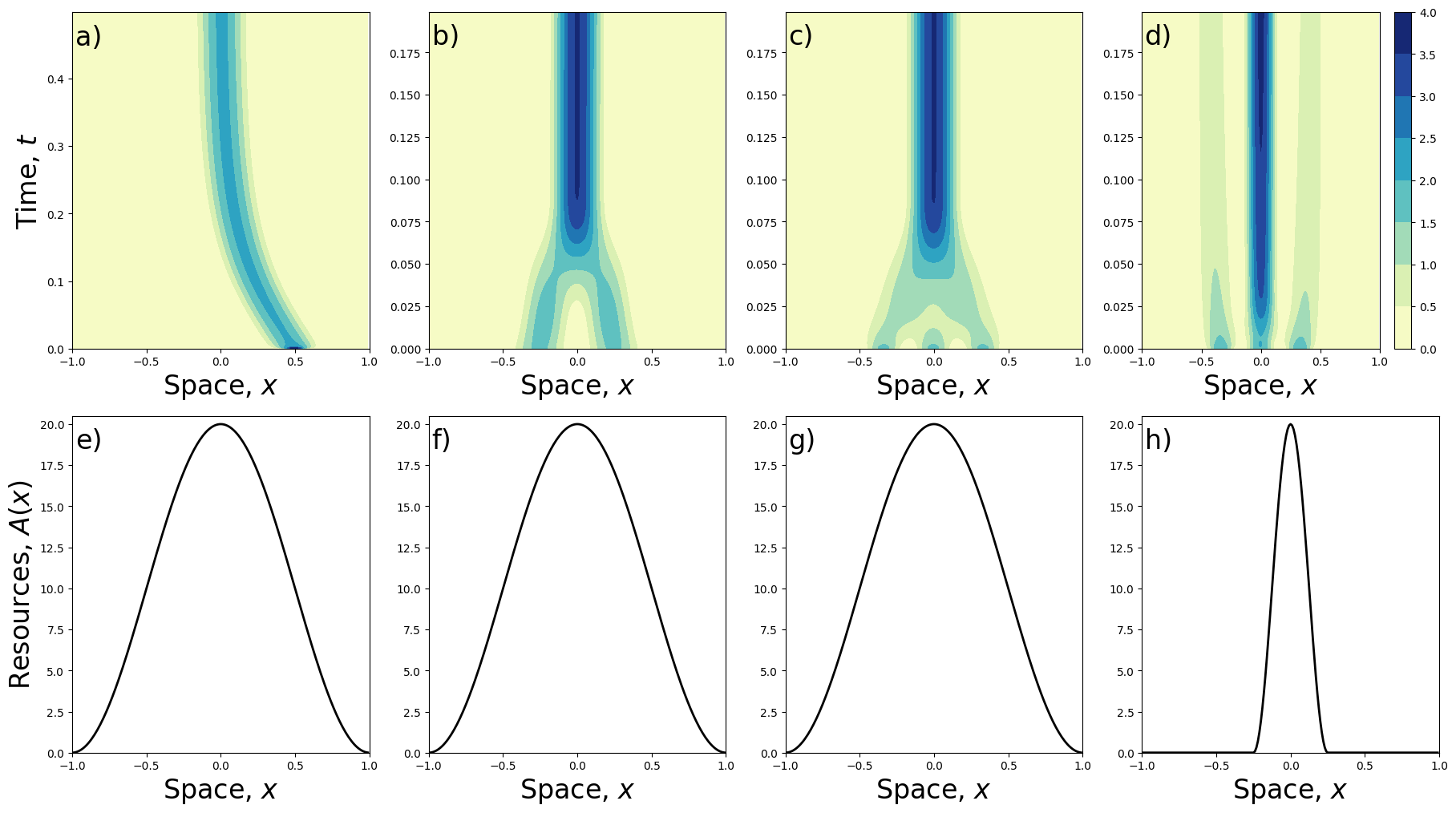}
	\end{center}
	\caption{{\bf The effect of initial conditions.} Panels (a-d) show numerical solutions to Equation (\ref{eq:adfond}) with $A(x)$ in the form of Equation (\ref{eq:ax_1hump}). Each solution has $\gamma=5$ and $\sigma=0.1$. The form of $A(x)$  for each numerical solution in Panels (a-d) is shown in Panels (e-h) respectively. Panels (e-g) have $n=1$ and $a_1=10$, whilst Panel (h) has $n=4$ and $a_4=10$, giving a narrower resource distribution.}
	\label{fig:diffic}
\end{figure} 
To address (i), Figure \ref{fig:diffic} displays numerical solutions to Equation (\ref{eq:adfond}) with a single clump of resources at the centre but with various different initial conditions. Specifically, the initial condition for Figure \ref{fig:diffic}a is a Gaussian distribution offset from the origin with mean $0.5$ and standard deviation $0.05$. 
Figure \ref{fig:diffic}b has initial condition
\begin{align}
	u(x,0)=\begin{cases}
		1-\cos\left(4\pi x\right), & \mbox{if $|x|<0.5$}, \\
		0, & \mbox{otherwise.}
	\end{cases}
\end{align}
Figure \ref{fig:diffic}c,d each have initial condition
\begin{align}
	u(x,0)=\begin{cases}
		1+\cos\left(6\pi x\right), & \mbox{if $|x|<0.5$}, \\
		0, & \mbox{otherwise.} 
	\end{cases}
\end{align}
The resource landscapes are given by Equation (\ref{eq:ax_1hump}), with $n=1$ and $a_1=10$ for Figures \ref{fig:diffic}a-c and $n=4$ and $a_4=10$ for Figure \ref{fig:diffic}d. These resource landscapes are displayed in Figures \ref{fig:diffic}e-h, respectively. 

Figure \ref{fig:diffic}a shows how a resource clump can cause an aggregation to move from its initial location towards the resource. Figures \ref{fig:diffic}b,c demonstrate the ability of a resource clump to unify multi-peaked initial conditions if it is sufficiently wide to capture all the peaks. However, as Figure \ref{fig:diffic}d shows, if the resource clump is too narrow, it may fail to suck in all the aggregations on the landscape. This demonstrates the importance of initial conditions in determining the final state of the system. Indeed, in Figure \ref{fig:diffic}d, the numerical algorithm has found a different local minimum energy to those discussed in Section \ref{sec:scl}, in which an assumption was made about the final distribution being single-peaked.

\begin{figure}[h!]
	\begin{center}
		\includegraphics[width=\columnwidth]{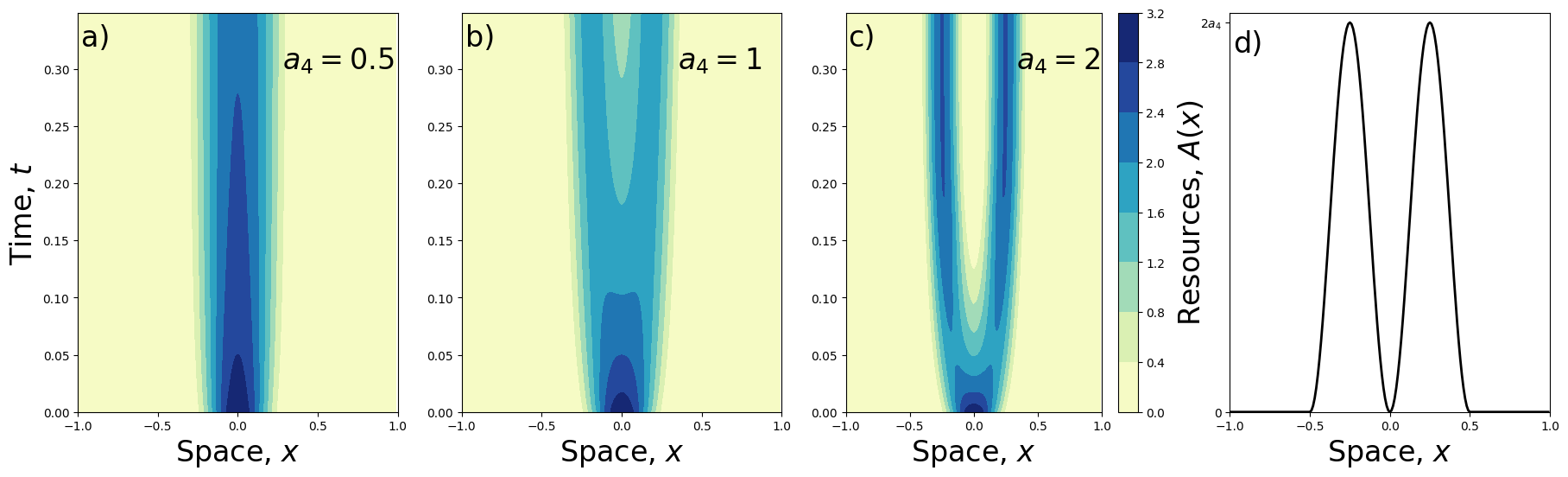}
	\end{center}
	\caption{{\bf Two-clumped resource layers.}  Panels (a-c) show numerical solutions to Equation (\ref{eq:adfond}) with $A(x)$ in the form of Equation (\ref{eq:ax_2clump}), given in Panel (d). The amount of the resources, given by $a_4$, affects the overall solution. Where $a_4$ is lower, as in Panel (a), the two clumps have the effect of stretching an  aggregation that is initially centred between the two clumps. For higher $a_4$, Panels (b) and (c) show that the resources can split apart the aggregations, with the extent of the splitting being more pronounced as $a_4$ is increased. For Panels (a-c), $\gamma=2$ and $\sigma=0.1$.}
	\label{fig:2clump}
\end{figure} 
To address (ii), Figure \ref{fig:2clump} shows solutions of Equation (\ref{eq:adfond}) with two clumps of attractive resources but an initial condition consisting of a single aggregation. The formula for the resource distribution is
\begin{align}
	\label{eq:ax_2clump}
A(x)=\begin{cases}
		a_4[1-\cos\left(4\pi x\right)], & \mbox{if $|x|<0.5$}, \\
		0, & \mbox{otherwise},
	\end{cases}
\end{align}
displayed in Figure \ref{fig:2clump}d. The initial condition for the numerical simulations is given by Equation (\ref{eq:ustarx_homr0}) (the minimum energy solution in the absence of resources). Figure \ref{fig:2clump} shows that small attraction to the two resource clumps ($a_4=0.5$) causes the aggregation to stretch slightly (Figure \ref{fig:2clump}a). For higher values of $a_4$, the resource attraction causes the aggregation to split in two (Figure \ref{fig:2clump}b), with the extent of the splitting appearing to increase for larger values of $a_4$ (Figure \ref{fig:2clump}c).

\section{Discussion}
\label{sec:disc}

This study analysed a PDE model of biological aggregation in a heterogeneous environment. This model consists of diffusion, non-local self-attraction (at the population level), and flow up the environmental gradient. These aspects of organism movement combine to shape emergent space use patterns, which are characterised as minimum energy solutions to the model. In the case where diffusion is quadratic and the non-local self-attraction is either through a Laplace kernel or a fourth-order approximation, analytic expressions are derived for the steady states of the model. When the environment consists of a single clump of attractive resources, finding the minimum energy steady state solution is quick and simple, compared to solving the PDE numerically. This enables some counter-intuitive results to be uncovered about how the environment interplays with self attraction: (a) a non-monotonic dependence of resource clump width on the organisms' aggregation width, and (b) a tendency for increased self-attraction to cause the emergent aggregation to increase in width, in situations where the resource attraction is very strong.

When making simplifications for the sake of analytic tractability, there is always a danger that the observed phenomena are just a feature of the simplified model and do not carry over to other, perhaps more realistic, modelling scenarios. Therefore our analysis is bolstered with numerical solutions of slightly-modified forms of our study PDE, focusing on the non-monotonic dependence mentioned in (a) above. This non-monotonic phenomenon seems to hold both when we examined a top-hat kernel rather than a fourth order approximation, and when we switched from quadratic to linear diffusion. Consequently, it seems that this phenomenon it is likely to be a genuine feature of biological aggregations and not just a quirk of a particular model formulation. Conversely, we could not find a situation where Case (b) held in situations away from the fourth-order approximation and quadratic diffusion, so this may just be an artefact of the particular model chosen.

It is worth making some conjectures about how the non-monotonic dependence from Case (a) might come about physically. It seems from the numerics that, as the clump width decreases, there is a transition from a simple up-and-down aggregation shape, to one where there is a clear wide part at the bottom and narrow part at the top (compare black solid and red dot-dash curves in Figure \ref{fig:adh_plot_num_ss}c). In other words, when the resource clump is relatively narrow (red dot-dash curves), some of the organisms follow the shape of the resource clump, but others cling on to the left and right of the central part of the aggregation in a manner governed by the width of the self-attraction kernel. However, when the resource clump is wider, this separation between a resource-induced narrow aggregation and a wider self-aggregation is no longer apparent, with these two features of aggregation seeming to `work together' to form the overall aggregative shape.
These patterning differences may help demarcate situations where there is evidence of both external features (e.g. chemical signals in the case of cell biology) and self-attraction combining to cause aggregative phenomena \citep{borgeretal2008, ellison2024combining, ho2019feather}.

The results presented here contribute to the general question of how to disentangle heterogeneity-induced patterns from  Turing-like instability-induced patterns. Early works in this direction include numerical investigation of a two-species reaction-diffusion system living amongst an immobile third  species with heterogeneous initial conditions \citep{voroney1996turing} and a two-species system where the interaction mechanism is spatially dependent \citep{cantrell1998effects}. This latter study involves heterogeneity in the reaction terms, an idea that has been further explored in various more recent studies of reaction-diffusion equations (e.g. \citet{krause2018heterogeneity, kozak2019pattern}), with others also including heterogeneity in the diffusion term \citep{van2021pattern}. The approach of the present work is slightly different as the equations do not involve reaction terms but instead have an advection term with spatial heterogeneity, alongside a nonlocal self-advection. There are various studies of pattern formation with heterogeneous advection alongside diffusion \citep{cantrell2008approximating,cantrell2010evolution}, but the inclusion of nonlocal self-advection with spatial heterogeneity in studies of pattern formation in PDEs appears to be less well understood.

This study provides the groundwork for understanding multi-species systems of diffusion and non-local advection in heterogeneous environments. Whilst the aggregation-diffusion equation is implicitly a single-species model, multi-species versions have gained much attention in recent years due to their applications in cell biology \citep{burger2018sorting}, ecology \citep{pottslewis2019}, and human behaviour \citep{barbaro2021analysis}. When the environment is homogeneous, regularity properties are becoming well-understood \citep{jungel2022nonlocal, giunta2022local, giunta2023positivity,carrillo2024well}, and there are several studies revealing rich patterning and bifurcation structures \citep{jewell2023patterning, giunta2024weakly, painter2024variations}. However, as in the single species situation, these mathematical models are motivated by biological systems that usually exist in heterogeneous environments, and these environments may qualitatively alter the emergent patterns. For example, the chase-and-run dynamics of predators and prey can be `pinned' by environmental heterogeneity, such as near the edge of a forest that prey use to hide from predators \citep{bonnot2013habitat}. Likewise, territorial animals may share space more in areas with abundant resources, leading to disparities in overlap driven by environmental heterogeneity \citep{sells2020economics}. Therefore it is important to move beyond the assumption of an homogeneous environment in both single- and multi-species models of non-local advection, to increase the biological relevance of these models and widen their scope of application.

\backmatter

\bmhead{Acknowledgements}

The author would like to thank the Faculty of Science at the University of Sheffield for granting him study leave, partly used for the research reported here, Dr. Andrew Krause for valuable discussions on the topic of this manuscript, and two anonymous reviewers who made helpful comments on a previous version of this manuscript.

\section*{Declarations}


\begin{itemize}
\item {\bf Funding.} No external funding was received for conducting this study.
\item {\bf Competing interests.} The author is an editorial board member of the Journal of Mathematical Biology, but has no other competing interests to declare. 
\item {\bf Data availability.} This manuscript has no associated data. 
\item {\bf Code availability.}  All code for performing the numerics in this study is available on GitHub at \url{https://github.com/jonathan-potts/AggDiffHet}. 
\end{itemize}


\clearpage
\begin{appendices}

\section{Alterations required for using a Laplace kernel}\label{sec:AppA}

Here, we detail how to modify the results of Section \ref{sec:scl} for analysing the Laplace kernel model from Equation (\ref{eq:aggdiffnd}), rather than the fourth-order model from Equation (\ref{eq:adfond}). The steady state solutions given in Equations (\ref{eq:ustarx_1hump})-(\ref{eq:1hump_Qcts}) are almost identical in the Laplace kernel model, except we need to replace $r_0$ with $r_1$, defined as
\begin{linenomath*}\begin{align}
		\label{eq:r1}
		r_1 = \sqrt{\frac{\pi^2}{m^2(\gamma-1)}}.
\end{align}\end{linenomath*}
Then
    \begin{linenomath*}\begin{align}
		\label{eq:ustarx_1humpl}
		u_\ast(x)=\begin{cases}\alpha_0+\alpha_n[1+\cos(\pi nx)]+Q\cos\left(\frac{x\pi}{r_1}\right), & \mbox{for $x \in \left[-\frac{1}{n},\frac{1}{n}\right]\cap[-r,r]$,}\\
			\alpha_0+Q\cos\left(\frac{x\pi}{r_1}\right), & \mbox{for $\frac{1}{n}<|x|\leq r$ if $r>\frac{1}{n}$,}\\
			0, &\mbox{otherwise,}\end{cases}
\end{align}\end{linenomath*}
and
\begin{linenomath*}\begin{align}
		\label{eq:1hump_alpha0l}
		\alpha_0=\begin{cases}\frac{p}{2r}-\alpha_n - \frac{\alpha_n}{\pi n r}\sin(\pi n r)-\frac{Qr_1}{\pi r}\sin\left(\frac{r\pi}{r_1}\right),&\mbox{if $r<1/n$,} \\
			\frac{p}{2r}-\frac{\alpha_n}{n r}-\frac{Qr_1}{\pi r}\sin\left(\frac{r\pi}{r_1}\right),&\mbox{otherwise,}
		\end{cases}
\end{align}\end{linenomath*}
and
\begin{linenomath*}\begin{align}
		\label{eq:1hump_Qctsl}
		Q = \begin{cases}-\frac{\alpha_0+\alpha_n+\alpha_n\cos(\pi n r)}{\cos(r\pi/r_1)}, & \mbox{if $r<1/n$,} \\
			-\frac{\alpha_0}{\cos(r\pi/r_1)},&\mbox{otherwise.}
		\end{cases}
\end{align}\end{linenomath*}
For the energy functional, analogous to Equation (\ref{eq:1hump_energy}), we calculate
\begin{linenomath*}\begin{align}
\label{eq:kmcos1}
K_m\ast\cos(n\pi x)&=\frac{m}{2}\int_{-\infty}^\infty {\rm e}^{-m|y|}\cos(n\pi(x+y)){\rm d}y \nonumber \\
&=\frac{m^2}{m^2+\pi^2 n^2}\cos(n\pi x).
\end{align}\end{linenomath*}
Similarly,
\begin{linenomath*}\begin{align}
\label{eq:kmsin1}
K_m\ast\sin(n\pi x)	&=\frac{m^2}{m^2+\pi^2 n^2}\sin(n\pi x), \\
\label{eq:kmcos2}
K_m\ast\cos\left(\frac{x\pi}{r_1}\right)&=\frac{1}{\gamma}\cos(n\pi x),  \\
\label{eq:kmsin2}
K_m\ast\sin\left(\frac{x\pi}{r_1}\right)&=\frac{1}{\gamma}\sin(n\pi x).
\end{align}\end{linenomath*}
Plugging Equations (\ref{eq:kmcos1})-(\ref{eq:kmsin2}) into Equation (\ref{eq:ustarx_1humpl}), then in turn plugging the result into Equation (\ref{eq:energy}) (the energy functional), a direct calculation leads (perhaps surprisingly) to exactly the same functional form as Equation (\ref{eq:1hump_energy}), but with $u_\ast(x)$ as given in Equation (\ref{eq:ustarx_1humpl}).

\section{Energy minimiser in an homogeneous landscape}\label{sec:AppB}

Here, we prove Equation (\ref{eq:ustarx_homr0}). This has been derived in \citet[Section 2.3.1]{falco2023local}, but we include our proof here as it is slightly different and uses notation consistent with the Main Text. We  assume that the support of $u_\ast(x)$ is a disjoint union of intervals.  Then, without loss of any further generality, we can assume further that the support is $(-r,r)$.  We are interested in the case $\gamma>1$, as this is the situation where the homogeneous steady state can be unstable to non-constant perturbations (the eigenvalue in this case is $\lambda =\kappa^2[(\gamma-1)-\kappa^2(\gamma\sigma^2/2)]p/2$ for wavenumber $\kappa$).

From Equation (\ref{eq:ustarx4})-(\ref{eq:ui4}), the general form of the steady state solution, given these constraints, is 
\begin{linenomath*}\begin{align}
		\label{eq:ustarx_hom}
		u_\ast(x)=\begin{cases}\alpha_0+Q\cos\left(\frac{x\pi}{r_0}\right), & \mbox{if $-r<x<r$,}\\
			0, &\mbox{otherwise,}\end{cases}
\end{align}\end{linenomath*}
for constants $\alpha_0$ and $Q$. Applying the integral condition from Equation (\ref{eq:p}), we find that 
\begin{linenomath*}\begin{align}
		\label{eq:alpha0_hom}
		\alpha_0 = \frac{p}{2r}-\frac{Qr_0}{\pi r}\sin\left(\frac{r\pi}{r_0}\right).
\end{align}\end{linenomath*}
\begin{proposition}\label{prop:hom}The case $r=r_0$ and $Q=\alpha_0$, given by
	\begin{linenomath*}\begin{align}
			\label{eq:ustarx_homr0_app}
			u_\ast(x)=\begin{cases}\frac{p}{2r_0}\left[1+\cos\left(\frac{x\pi}{r_0}\right)\right], & \mbox{if $-r_0<x<r_0$,}\\
				0, &\mbox{otherwise,}\end{cases}
	\end{align}\end{linenomath*}
	minimises the energy locally amongst continuous, positive, steady-state solutions.
\end{proposition}
\noindent{\bf Proof.} First, a direct calculation shows that the energy when $r=r_0$ is
\begin{linenomath*}\begin{align}
		E[u_\ast]_{r=r_0}=\frac{(1-\gamma)p^2}{2r_0}.
\end{align}\end{linenomath*}
Let $\epsilon>0$ be arbitrarily small.  We only need to examine the case $r=r_0-\epsilon$, as there is no solution that is both continuous and positive when $r=r_0+\epsilon$.  We can then calculate directly (but with some effort) to find
\begin{linenomath*}\begin{align}
		E[u_\ast]_{r=r_0-\epsilon}=(\gamma-1)\left[\frac{Qr_0p}{\pi r}\sin\left(\frac{\epsilon\pi}{r_0}\right)-\frac{p^2}{2r}\right].
\end{align}\end{linenomath*}
Expanding this to first order in $\epsilon$ gives
\begin{linenomath*}\begin{align}
		\label{eq:energy_exp}
		E[u_\ast]_{r=r_0-\epsilon}=(\gamma-1)\left[-\frac{p^2}{2r_0}+\left(\frac{Qp}{r_0}-\frac{p^2}{2r_0^2}\right)\epsilon\right]+O(\epsilon^2).
\end{align}\end{linenomath*}
For $u_\ast(x)$ to be continuous, we require $u_\ast(r)=0$ so that 
\begin{linenomath*}\begin{align}
		Q=p\left[\frac{2r_0}{\pi }\sin\left(\frac{\epsilon\pi}{r_0}\right)-2r\cos\left(\frac{r\pi}{r_0}\right)\right]^{-1}.
\end{align}\end{linenomath*}
Expanding the term inside the square brackets to second order in $\epsilon$, we find
\begin{linenomath*}\begin{align}
		\frac{2r_0}{\pi }\sin\left(\frac{\epsilon\pi}{r_0}\right)-2r\cos\left(\frac{r\pi}{r_0}\right) \approx 2r_0-\frac{\pi^2\epsilon^2}{r_0}<2r_0.
\end{align}\end{linenomath*}
Hence $Q>p/2r_0$ so
\begin{linenomath*}\begin{align}
		\frac{Qp}{r_0}-\frac{p^2}{2r_0^2}>0,
\end{align}\end{linenomath*}
and therefore, by Equation (\ref{eq:energy_exp}), we have 
\begin{linenomath*}\begin{align}
		E[u_\ast]_{r=r_0-\epsilon}>\frac{(1-\gamma)p^2}{2r_0}=	E[u_\ast]_{r=r_0},
\end{align}\end{linenomath*}
for arbitrarily small $\epsilon$.   
\qed

\section{Propositions \ref{thm:ss1} and \ref{thm:ss2} for $\gamma=1$}\label{sec:AppC}

To find an expression analogous to Equations (\ref{eq:ustarx}-\ref{eq:ui}) with $\gamma=1$, note that Equation (\ref{eq:aggdiff_ss}) becomes
\begin{linenomath*}\begin{align}
		\label{eq:aggdiff_ss_app}
		\frac{{\rm d}^2 u_*}{{\rm d} x^2}=\frac{{\rm d}^2 A}{{\rm d} x^2}-m^2c-m^2A(x),
\end{align}\end{linenomath*}
which can be integrated twice to give
\begin{linenomath*}\begin{align}
		\label{eq:aggdiff_ss_app2}
		u_*(x)=A(x)-m^2cx^2-m^2B(x),
\end{align}\end{linenomath*}
where 
\begin{linenomath*}\begin{align}
		\label{eq:axnd_app}
		B(x)=P+Qx+\frac{a_0}{2}x^2-\sum_{n=1}^\infty\frac{1}{n^2\pi^2}\left[a_n\cos\left({n\pi x}\right)+b_n\sin\left({n\pi x}\right)\right],
\end{align}\end{linenomath*}
and $P,Q \in {\mathbb R}$ are arbitrary constants.

Similarly, for an expression analogous to Equations (\ref{eq:ustarx4}--\ref{eq:ui4}) with $\gamma=1$, note that Equation (\ref{eq:adfo_ssz2}) becomes
\begin{linenomath*}\begin{align}
		\label{eq:adfo_ssz2_app}
		\frac{{\rm d}^2 u_\ast}{{\rm d} x^2}=-\frac{2}{\sigma^2}(c+A),
\end{align}\end{linenomath*}
which can be integrated twice to give
\begin{linenomath*}\begin{align}
		\label{eq:adfo2_ssz2_app}
		u_*(x)=-\frac{2}{\sigma^2}(cx^2+B(x)).
\end{align}\end{linenomath*}

\end{appendices}

\bibliography{adh_refs}

\begin{thebibliography}{53}
\providecommand{\natexlab}[1]{#1}
\providecommand{\url}[1]{{#1}}
\providecommand{\urlprefix}{URL }
\providecommand{\doi}[1]{\url{https://doi.org/#1}}
\providecommand{\eprint}[2][]{\url{#2}}
 \bibcommenthead

\bibitem[{Aarts et~al(2008)Aarts, MacKenzie, McConnell, Fedak, and
  Matthiopoulos}]{aarts2008estimating}
Aarts G, MacKenzie M, McConnell B, et~al (2008) Estimating space-use and
  habitat preference from wildlife telemetry data. Ecography 31(1):140--160

\bibitem[{Bailo et~al(2020)Bailo, Carrillo, and Hu}]{bailo2020fully}
Bailo R, Carrillo JA, Hu J (2020) Fully discrete positivity-preserving and
  energy-dissipating schemes for aggregation-diffusion equations with a
  gradient-flow structure. Communications in Mathematical Sciences
  18(5):1259--1303

\bibitem[{Bailo et~al(2023)Bailo, Carrillo, Kalliadasis, and
  Perez}]{bailo2023unconditional}
Bailo R, Carrillo J, Kalliadasis S, et~al (2023) Unconditional bound-preserving
  and energy-dissipating finite-volume schemes for the cahn-hilliard equation.
  Communications in Computational Physics 34

\bibitem[{Barbaro et~al(2021)Barbaro, Rodriguez, Yolda{\c{s}}, and
  Zamponi}]{barbaro2021analysis}
Barbaro AB, Rodriguez N, Yolda{\c{s}} H, et~al (2021) Analysis of a
  cross-diffusion model for rival gangs interaction in a city. Communications
  in Mathematical Sciences 19(8):2139--2175

\bibitem[{Bastille-Rousseau et~al(2018)Bastille-Rousseau, Douglas-Hamilton,
  Blake, Northrup, and Wittemyer}]{bastille2018applying}
Bastille-Rousseau G, Douglas-Hamilton I, Blake S, et~al (2018) Applying network
  theory to animal movements to identify properties of landscape space use.
  Ecological Applications 28(3):854--864

\bibitem[{Bonner(2009)}]{bonner2009}
Bonner JT (2009) The social amoebae: the biology of cellular slime molds.
  Princeton University Press

\bibitem[{Bonnot et~al(2013)Bonnot, Morellet, Verheyden, Cargnelutti, Lourtet,
  Klein, and Hewison}]{bonnot2013habitat}
Bonnot N, Morellet N, Verheyden H, et~al (2013) Habitat use under predation
  risk: hunting, roads and human dwellings influence the spatial behaviour of
  roe deer. European journal of wildlife research 59:185--193

\bibitem[{B\"orger et~al(2008)B\"orger, Dalziel, and Fryxell}]{borgeretal2008}
B\"orger L, Dalziel BD, Fryxell JM (2008) {Are there general mechanisms of
  animal home range behaviour? A review and prospects for future research}.
  Ecol Lett 11(6):637--650. \doi{10.1111/j.1461-0248.2008.01182.x},
  \urlprefix\url{http://dx.doi.org/10.1111/j.1461-0248.2008.01182.x}

\bibitem[{Boyce et~al(2016)Boyce, Johnson, Merrill, Nielsen, Solberg, and
  Van~Moorter}]{boyce2016can}
Boyce MS, Johnson CJ, Merrill EH, et~al (2016) Can habitat selection predict
  abundance? Journal of Animal Ecology 85(1):11--20

\bibitem[{Briscoe et~al(2002)Briscoe, Lewis, and Parrish}]{briscoeetal2002}
Briscoe B, Lewis M, Parrish S (2002) Home range formation in wolves due to
  scent marking. Bull Math Biol 64(2):261--284. \doi{10.1006/bulm.2001.0273},
  \urlprefix\url{http://dx.doi.org/10.1006/bulm.2001.0273}

\bibitem[{Burger et~al(2018)Burger, Francesco, Fagioli, and
  Stevens}]{burger2018sorting}
Burger M, Francesco MD, Fagioli S, et~al (2018) Sorting phenomena in a
  mathematical model for two mutually attracting/repelling species. SIAM
  Journal on Mathematical Analysis 50(3):3210--3250

\bibitem[{Cantrell and Cosner(1998)}]{cantrell1998effects}
Cantrell RS, Cosner C (1998) On the effects of spatial heterogeneity on the
  persistence of interacting species. Journal of Mathematical Biology
  37:103--145

\bibitem[{Cantrell et~al(2008)Cantrell, Cosner, and
  Lou}]{cantrell2008approximating}
Cantrell RS, Cosner C, Lou Y (2008) Approximating the ideal free distribution
  via reaction--diffusion--advection equations. Journal of Differential
  Equations 245(12):3687--3703

\bibitem[{Cantrell et~al(2010)Cantrell, Cosner, and
  Lou}]{cantrell2010evolution}
Cantrell RS, Cosner C, Lou Y (2010) Evolution of dispersal and the ideal free
  distribution. Mathematical Biosciences \& Engineering 7(1):17--36

\bibitem[{Carrillo et~al(2019)Carrillo, Craig, and
  Yao}]{carrillo2019aggregation}
Carrillo JA, Craig K, Yao Y (2019) Aggregation-diffusion equations: dynamics,
  asymptotics, and singular limits. In: Active Particles, Volume 2. Springer, p
  65--108

\bibitem[{Carrillo et~al(2024)Carrillo, Salmaniw, and
  Skrzeczkowski}]{carrillo2024well}
Carrillo JA, Salmaniw Y, Skrzeczkowski J (2024) Well-posedness of
  aggregation-diffusion systems with irregular kernels. arXiv preprint
  arXiv:240609227

\bibitem[{Chen et~al(2015)Chen, Foley, Tang, Li, Jiang, Wu, Widelitz, and
  Chuong}]{chen2015development}
Chen CF, Foley J, Tang PC, et~al (2015) Development, regeneration, and
  evolution of feathers. Annu Rev Anim Biosci 3(1):169--195

\bibitem[{Ellison et~al(2024)Ellison, Potts, Strickland, Demarais, and
  Street}]{ellison2024combining}
Ellison N, Potts JR, Strickland BK, et~al (2024) Combining animal interactions
  and habitat selection into models of space use: a case study with
  white-tailed deer. Wildlife Biology 2024(3):e01,211

\bibitem[{Fagan et~al(2017)Fagan, Gurarie, Bewick, Howard, Cantrell, and
  Cosner}]{fagan2017perceptual}
Fagan WF, Gurarie E, Bewick S, et~al (2017) Perceptual ranges, information
  gathering, and foraging success in dynamic landscapes. The American
  Naturalist 189(5):474--489

\bibitem[{Falc{\'o} et~al(2023)Falc{\'o}, Baker, and Carrillo}]{falco2023local}
Falc{\'o} C, Baker RE, Carrillo JA (2023) A local continuum model of cell-cell
  adhesion. SIAM Journal on Applied Mathematics pp S17--S42

\bibitem[{Georgiou et~al(2021)Georgiou, Buhl, Green, Lamichhane, and
  Thamwattana}]{georgiou2021modelling}
Georgiou F, Buhl J, Green JEF, et~al (2021) Modelling locust foraging: How and
  why food affects group formation. PLOS Computational Biology 17(7):e1008,353

\bibitem[{Giunta et~al(2022{\natexlab{a}})Giunta, Hillen, Lewis, and
  Potts}]{giunta2022local}
Giunta V, Hillen T, Lewis M, et~al (2022{\natexlab{a}}) Local and global
  existence for nonlocal multispecies advection-diffusion models. SIAM Journal
  on Applied Dynamical Systems 21(3):1686--1708

\bibitem[{Giunta et~al(2022{\natexlab{b}})Giunta, Hillen, Lewis, and
  Potts}]{giunta2022detecting}
Giunta V, Hillen T, Lewis MA, et~al (2022{\natexlab{b}}) Detecting minimum
  energy states and multi-stability in nonlocal advection--diffusion models for
  interacting species. Journal of Mathematical Biology 85(5):56

\bibitem[{Giunta et~al(2023)Giunta, Hillen, Lewis, and
  Potts}]{giunta2023positivity}
Giunta V, Hillen T, Lewis M, et~al (2023) Positivity and global existence for
  nonlocal advection-diffusion models of interacting populations. arXiv
  preprint arXiv:231209692

\bibitem[{Giunta et~al(2024)Giunta, Hillen, Lewis, and
  Potts}]{giunta2024weakly}
Giunta V, Hillen T, Lewis MA, et~al (2024) Weakly nonlinear analysis of a
  two-species non-local advection--diffusion system. Nonlinear Analysis: Real
  World Applications 78:104,086

\bibitem[{Hillen and Painter(2013)}]{hillenpainter2013}
Hillen T, Painter K (2013) Transport and anisotropic diffusion models for
  movement in oriented habitats. In: Lewis MA, Maini PK, Petrovskii SV (eds)
  Dispersal, Individual Movement and Spatial Ecology. Lecture Notes in
  Mathematics, Springer Berlin Heidelberg, p 177--222,
  \doi{10.1007/978-3-642-35497-7_7},
  \urlprefix\url{http://dx.doi.org/10.1007/978-3-642-35497-7_7}

\bibitem[{Ho et~al(2019)Ho, Freem, Zhao, Painter, Woolley, Gaffney, McGrew,
  Tzika, Milinkovitch, Schneider et~al}]{ho2019feather}
Ho WK, Freem L, Zhao D, et~al (2019) Feather arrays are patterned by
  interacting signalling and cell density waves. PLoS Biology 17(2):e3000,132

\bibitem[{Horne et~al(2008)Horne, Garton, and Rachlow}]{horne2008synoptic}
Horne JS, Garton EO, Rachlow JL (2008) A synoptic model of animal space use:
  simultaneous estimation of home range, habitat selection, and
  inter/intra-specific relationships. ecological modelling 214(2-4):338--348

\bibitem[{Hueschen et~al(2023)Hueschen, Dunn, and
  Phillips}]{hueschen2023wildebeest}
Hueschen CL, Dunn AR, Phillips R (2023) Wildebeest herds on rolling hills:
  Flocking on arbitrary curved surfaces. Physical Review E 108(2):024,610

\bibitem[{Jewell et~al(2023)Jewell, Krause, Maini, and
  Gaffney}]{jewell2023patterning}
Jewell TJ, Krause AL, Maini PK, et~al (2023) Patterning of nonlocal transport
  models in biology: the impact of spatial dimension. Mathematical Biosciences
  366:109,093

\bibitem[{J{\"u}ngel et~al(2022)J{\"u}ngel, Portisch, and
  Zurek}]{jungel2022nonlocal}
J{\"u}ngel A, Portisch S, Zurek A (2022) Nonlocal cross-diffusion systems for
  multi-species populations and networks. Nonlinear Analysis 219:112,800

\bibitem[{Kim et~al(2016)Kim, Lee, Choi, Lee, and Jeong}]{kim2016basic}
Kim J, Lee S, Choi Y, et~al (2016) Basic principles and practical applications
  of the cahn--hilliard equation. Mathematical Problems in Engineering
  2016(1):9532,608

\bibitem[{Koz{\'a}k et~al(2019)Koz{\'a}k, Gaffney, and
  Klika}]{kozak2019pattern}
Koz{\'a}k M, Gaffney EA, Klika V (2019) Pattern formation in reaction-diffusion
  systems with piecewise kinetic modulation: an example study of heterogeneous
  kinetics. Physical Review E 100(4):042,220

\bibitem[{Krause et~al(2018)Krause, Klika, Woolley, and
  Gaffney}]{krause2018heterogeneity}
Krause AL, Klika V, Woolley TE, et~al (2018) Heterogeneity induces
  spatiotemporal oscillations in reaction-diffusion systems. Physical Review E
  97(5):052,206

\bibitem[{Morales et~al(2021)Morales, Raspopovic, and
  Marcon}]{morales2021embryos}
Morales JS, Raspopovic J, Marcon L (2021) From embryos to embryoids: How
  external signals and self-organization drive embryonic development. Stem Cell
  Reports 16(5):1039--1050

\bibitem[{Painter et~al(2024{\natexlab{a}})Painter, Giunta, Potts, and
  Bernardi}]{painter2024variations}
Painter KJ, Giunta V, Potts JR, et~al (2024{\natexlab{a}}) Variations in
  non-local interaction range lead to emergent chase-and-run in heterogeneous
  populations. Journal of the Royal Society Interface 21(219):20240,409

\bibitem[{Painter et~al(2024{\natexlab{b}})Painter, Hillen, and
  Potts}]{painter2024biological}
Painter KJ, Hillen T, Potts JR (2024{\natexlab{b}}) Biological modeling with
  nonlocal advection--diffusion equations. Mathematical Models and Methods in
  Applied Sciences 34(01):57--107

\bibitem[{Papadopoulou et~al(2022)Papadopoulou, Hildenbrandt, Sankey, Portugal,
  and Hemelrijk}]{papadopoulou2022self}
Papadopoulou M, Hildenbrandt H, Sankey DW, et~al (2022) Self-organization of
  collective escape in pigeon flocks. PLoS Computational Biology
  18(1):e1009,772

\bibitem[{Patlak(1953)}]{patlak1953random}
Patlak CS (1953) Random walk with persistence and external bias. The Bulletin
  of Mathematical Biophysics 15:311--338

\bibitem[{Potts and B{\"o}rger(2023)}]{potts2023scale}
Potts JR, B{\"o}rger L (2023) How to scale up from animal movement decisions to
  spatiotemporal patterns: An approach via step selection. Journal of Animal
  Ecology 92(1):16--29

\bibitem[{Potts and Lewis(2016)}]{pottslewis2016b}
Potts JR, Lewis MA (2016) Territorial pattern formation in the absence of an
  attractive potential. J Math Biol 72(1-2):25--46

\bibitem[{Potts and Lewis(2019)}]{pottslewis2019}
Potts JR, Lewis MA (2019) Spatial memory and taxis-driven pattern formation in
  model ecosystems. Bulletin of Mathematical Biology 81(7):2725--2747.
  \doi{10.1007/s11538-019-00626-9},
  \urlprefix\url{https://doi.org/10.1007/s11538-019-00626-9}

\bibitem[{Potts and Schl{\"a}gel(2020)}]{potts2020parametrizing}
Potts JR, Schl{\"a}gel UE (2020) Parametrizing diffusion-taxis equations from
  animal movement trajectories using step selection analysis. Methods in
  Ecology and Evolution 11(9):1092--1105

\bibitem[{Roussi(2020)}]{roussi2020}
Roussi A (2020) Why gigantic locust swarms are challenging governments and
  researchers. Nature 579(7798):330--331

\bibitem[{Sells and Mitchell(2020)}]{sells2020economics}
Sells SN, Mitchell MS (2020) The economics of territory selection. Ecological
  Modelling 438:109,329

\bibitem[{Stears et~al(2020)Stears, Schmitt, Wilmers, and
  Shrader}]{stears2020mixed}
Stears K, Schmitt MH, Wilmers CC, et~al (2020) Mixed-species herding levels the
  landscape of fear. Proceedings of the Royal Society B 287(1922):20192,555

\bibitem[{Strandburg-Peshkin et~al(2017)Strandburg-Peshkin, Farine, Crofoot,
  and Couzin}]{strandburg2017habitat}
Strandburg-Peshkin A, Farine DR, Crofoot MC, et~al (2017) Habitat and social
  factors shape individual decisions and emergent group structure during baboon
  collective movement. elife 6:e19,505

\bibitem[{Turchin(1989)}]{turchin1989}
Turchin P (1989) Population consequences of aggregative movement. Journal of
  Animal Ecology pp 75--100

\bibitem[{Van~Gorder(2021)}]{van2021pattern}
Van~Gorder RA (2021) Pattern formation from spatially heterogeneous
  reaction--diffusion systems. Philosophical Transactions of the Royal Society
  A 379(2213):20210,001

\bibitem[{Van~Moorter et~al(2016)Van~Moorter, Rolandsen, Basille, and
  Gaillard}]{van2016movement}
Van~Moorter B, Rolandsen CM, Basille M, et~al (2016) Movement is the glue
  connecting home ranges and habitat selection. Journal of Animal Ecology
  85(1):21--31

\bibitem[{Voroney et~al(1996)Voroney, Lawniczak, and
  Kapral}]{voroney1996turing}
Voroney JP, Lawniczak A, Kapral R (1996) Turing pattern formation in
  heterogenous media. Physica D: Nonlinear Phenomena 99(2-3):303--317

\bibitem[{Wang and Salmaniw(2023)}]{wang2023open}
Wang H, Salmaniw Y (2023) Open problems in pde models for knowledge-based
  animal movement via nonlocal perception and cognitive mapping. Journal of
  Mathematical Biology 86(5):71

\bibitem[{Widelitz et~al(2003)Widelitz, Jiang, Yu, Shen, Shen, Wu, Yu, and
  Chuong}]{widelitz2003molecular}
Widelitz RB, Jiang TX, Yu M, et~al (2003) Molecular biology of feather
  morphogenesis: A testable model for evo-devo research. Journal of
  Experimental Zoology Part B: Molecular and Developmental Evolution
  298(1):109--122

\end{thebibliography}

\end{document}